\PassOptionsToPackage{table,xcdraw}{xcolor}

\PassOptionsToPackage{per-mode=symbol}{siunitx}

\documentclass[nonacm, acmtog, balance=false]{acmart}
\usepackage{algpseudocode}
\usepackage{algorithm}
\usepackage[free-standing-units=true]{siunitx}
\usepackage{booktabs}
\usepackage{multirow}
\usepackage{arydshln}
\usepackage{siunitx}
\usepackage{overpic}
\usepackage{graphicx}
\usepackage{subcaption}
\usepackage{tikz}

\usetikzlibrary{calc}

\begin{document}

\title{Pitfalls of Projection: A study of Newton-type solvers for incremental potentials}
\author{Andreas Longva}
\affiliation{
	\institution{RWTH Aachen University}
	\country{Germany}
}
\author{Fabian Löschner}
\affiliation{
	\institution{RWTH Aachen University}
	\country{Germany}
}
\author{José Antonio Fernández-Fernández}
\affiliation{
	\institution{RWTH Aachen University}
	\country{Germany}
}
\date{}
\author{Egor Larionov}
\affiliation{
	\institution{Meta Reality Labs}
	\country{United States}
}
\author{Uri M. Ascher}
\affiliation{
	\institution{University of British Columbia}
	\country{Canada}
}
\author{Jan Bender}
\affiliation{
	\institution{RWTH Aachen University}
	\country{Germany}
}

\authorsaddresses{}

\renewcommand{\vec}[1]{\mathbf{\boldsymbol{#1}}}

\newcommand{\labeledref}[2]{\hyperref[#2]{#1~\ref{#2}}}

\newcommand{\sectionref}[1]{\labeledref{Section}{#1}}
\newcommand{\appendixref}[1]{\labeledref{Appendix}{#1}}
\newcommand{\algoref}[1]{\labeledref{Algorithm}{#1}}
\newcommand{\figref}[1]{\labeledref{Figure}{#1}}
\newcommand{\tableref}[1]{\labeledref{Table}{#1}}

\renewcommand{\d}[1]{\, \mathrm{d} #1}
\newcommand{\pd}[2]{\frac{\partial #1}{\partial #2}}
\newcommand{\pdd}[3]{\frac{\partial^2 #1}{\partial #2 \, \partial #3}}
\newcommand{\hessian}[2]{\frac{\partial^2 #1}{\partial #2^2}}

\newcommand{\norm}[1]{\| #1 \|}

\newcommand{\td}[2]{\frac{\mathrm{d} #1}{\mathrm{d} #2}}

\newcommand{\note}[1]{{\color{blue} #1}}
\newcommand{\todo}[1]{\note{TODO: #1}}

\newcommand\fl[1]{\textcolor{orange}{\textbf{FL: #1}}}
\newcommand\jf[1]{\textcolor{green!60!black}{\textbf{JF: #1}}}
\newcommand\al[1]{\textcolor{magenta}{\textbf{AL: #1}}}
\newcommand\lw[1]{\textcolor{cyan}{\textbf{LW: #1}}}
\newcommand\jb[1]{\textcolor[rgb]{0.8,0.0,0.0}{JB: #1}}
\newcommand\sj[1]{\textcolor{blue}{\textbf{SJ: #1}}}

\definecolor{table_alternate}{HTML}{F6F7FB}

\definecolor{commentcolor}{HTML}{306E10}
\renewcommand{\algorithmiccomment}[1]{\hfill $\triangleright$ \textcolor{commentcolor}{#1}}

\begin{abstract}

Nonlinear systems arising from time integrators like Backward Euler can sometimes be reformulated as optimization problems, known as incremental potentials.
We show through a comprehensive experimental analysis that the widely used \emph{Projected Newton} method, which relies on unconditional semidefinite projection of Hessian contributions, \emph{typically} exhibits a reduced convergence rate compared to classical Newton's method.
We demonstrate how factors like resolution, element order, projection method, material model and boundary handling impact convergence of Projected Newton and Newton.

Drawing on these findings, we propose the hybrid method \emph{Project-on-Demand Newton}, which projects only \emph{conditionally}, and show that it enjoys both the robustness of Projected Newton and convergence rate of Newton.
We additionally introduce \emph{Kinetic Newton}, a regularization-based method that takes advantage of the structure of incremental potentials and avoids projection altogether.
We compare the four solvers on hyperelasticity and contact problems.

We also present a nuanced discussion of convergence criteria, and propose a new acceleration-based criterion that avoids problems associated with existing residual norm criteria and is easier to interpret.
We finally address a fundamental limitation of the Armijo backtracking line search that occasionally blocks convergence, especially for stiff problems.
We propose a novel parameter-free, robust line search technique to eliminate this issue.

\end{abstract}
\newlength{\originalcolumnwidth}
\setlength{\originalcolumnwidth}{243.14749pt}

\begin{teaserfigure}
	\begin{minipage}{\originalcolumnwidth}
		\centering
		\includegraphics[width=\originalcolumnwidth, trim={2cm, 0, 2cm, 4cm}, clip]{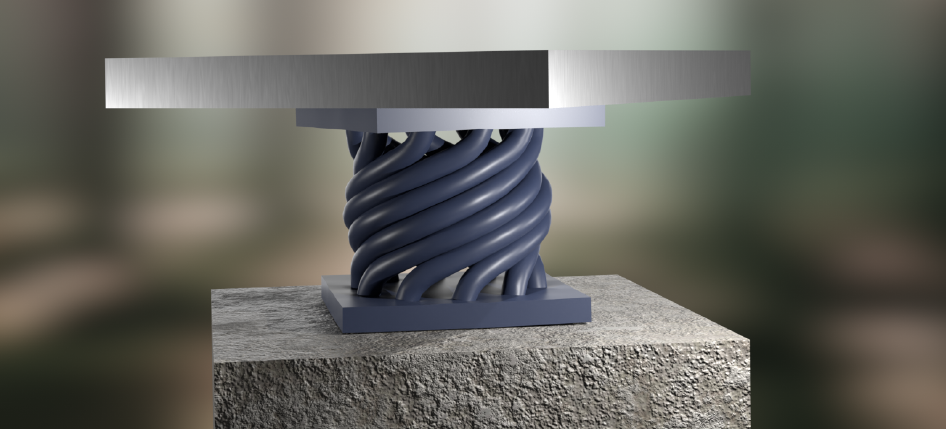}
	\end{minipage}
	\hfill
	\begin{minipage}{\originalcolumnwidth}
		\centering
	\begin{overpic}{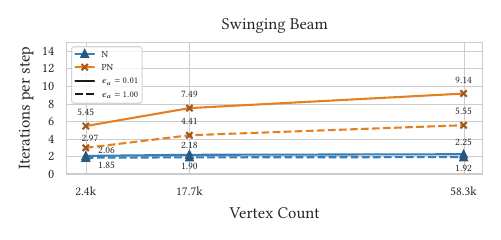}
		\put(40,19){\includegraphics[width=90pt]{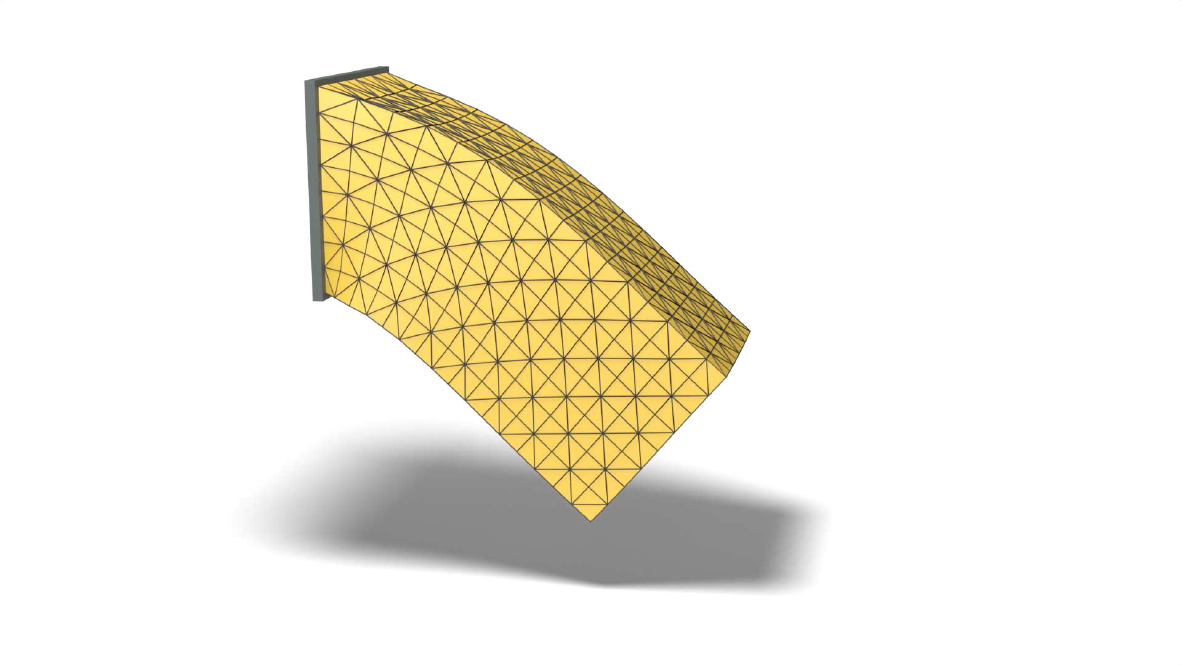}}
	\end{overpic}
	\end{minipage}
	\caption{
		\textbf{Left}: A rubber-like structure is twisted by the friction forces that arise as a rotating steel plate presses down upon it.
		This scenario is one of a series of benchmarks used to evaluate the Newton-type methods in this paper.
		\textbf{Right}: Average solver iterations per time step for the simulation of a swinging beam with the Newton (N) and Projected Newton (PN) methods, for coarse and fine parameters of the convergence tolerance, and for different resolutions.
		The figure overlaid on top depicts the lowest resolution simulation at the point of maximum deflection.
	    See \sectionref{sec:projected_newton_convergence} for more details on the experiment.
	}
	\label{fig:teaser}
\end{teaserfigure}

\maketitle

\section{Introduction}
Optimization-based time integrators are ubiquitous for simulation of deformable solids, multibody contact, cloth and many other time-dependent physical systems in computer graphics.
They rely on an \emph{Incremental Potential} (IP) reformulation of the discrete equations of motion, and the resulting optimization problems may be solved by Newton-type methods.
Most often, \emph{semidefinite projection}~\cite{TSI+05} is used to ensure positive definiteness of the system matrix, leading to the widely used \emph{Projected Newton} method (PN).

Despite its popularity in the graphics community, Projected Newton does not inherit key advantages of Newton's method.
It does not attain the same fast convergence rate, and does not enjoy affine invariance.
These facts seem to be relatively unknown, and Projected Newton appears to have escaped careful scrutiny.

To demonstrate that these concerns are not merely theoretical, we consider a trivial example in which a simple beam is clamped on one end and swinging under the influence of gravity.
We simulate the beam with Newton's method and Projected Newton for different resolutions and convergence tolerances.
We observe in \figref{fig:teaser} (Right) that Projected Newton needs far more iterations than Newton's method, and the discrepancy increases as the resolution of the beam increases.
We will study this example more in detail in \sectionref{sec:projected_newton_convergence},
and in \sectionref{sec:evaluation} we will see that this example is not an outlier; rather, the slower convergence is \emph{typical} of Projected Newton.

While Newton's method is superior for this particular example, this is often not the case.
If the energy that we try to minimize is not well-behaved, and the Hessian is not positive definite, Newton's method may make slow progress when far from the solution.
In this regime, Projected Newton excels.
This immediately suggests a strategy in which projection is not applied \emph{unconditionally}, but instead only when most effective.
To design such a method, we must first understand the effect semidefinite projection has on convergence.

To the best of our knowledge, we conduct the first experimental analysis of the effects semidefinite projection has in response to a number of factors, including mesh resolution, element order, projection method, material model, and boundary handling method, using classic Newton as a baseline for comparison.

Drawing on these insights, we design in \sectionref{sec:projection_on_demand} the simple hybrid method \emph{Project-on-Demand Newton} (POD-Newton).
By projecting only under certain conditions, it inherits Newton's local convergence rate,  significantly outperforming Projected Newton across all of our benchmarks,
and rapidly converging where Newton struggles.

We also explore a novel concept for adaptive regularization of the Hessian that circumvents the need for semidefinite projection altogether.
We show that with an appropriate choice of norm, the regularized Hessian matrix strongly resembles the Hessian for the same problem with a smaller time step, which enables us to design a simple and intuitive scheme for choosing the regularization parameter.
The resulting method, \emph{Kinetic Newton} (\sectionref{sec:kinetic_newton}), performs particularly well on smooth elasticity problems.

We consider each of the four methods to be a representative of its own class of Newton-type methods:
\begin{itemize}
\item \emph{No modification}: Classical Newton (baseline).
\item \emph{Unconditional projection}: Projected Newton.
\item \emph{Conditional projection}: Project-on-Demand Newton.
\item \emph{Regularization}: Kinetic Newton.
\end{itemize}
We evaluate all four methods on a set of benchmarks, ranging from academic examples of hyperelasticity to multibody contact with strong forces (\sectionref{sec:evaluation}), which gives us insights into the advantages and limitations of each class (\sectionref{sec:evaluation_discussion}).

While designing the experiments for our evaluation, we encountered two auxiliary challenges that complicated the process.

First, we found it difficult to consistently decide on convergence criteria for disparate experiments, or for a family of similar experiments where parameters are varied.
Different choices of convergence criterion make different trade-offs.
In \sectionref{sec:convergence_criteria} we discuss various options in order to contribute vital nuance to the literature.
We additionally propose a new criterion based on the \emph{balance of acceleration} that we believe is easier to reason about than residual norm tolerances.

Second, the backtracking line search would sometimes fail for stiff problems, irrespective of the solver. This made it difficult to fairly compare different solvers, as they would sometimes fail for reasons completely unrelated to the solver itself.
The cause is a fundamental limitation in the numerical computation of the Armijo condition, leading it to sometimes reject perfectly valid steps (\sectionref{sec:robust_line_search}).
We finally propose a new robust descent condition for the line search that, unlike previous work, is entirely parameter-free.
\section{Related work}
\paragraph{Incremental Potentials and Optimization-based Time Integrators}
\citet{RO99} recognized that the nonlinear equations associated with each time increment of integrators like Backward Euler may --- under certain assumptions --- be formulated as a potential minimization problem, called an \emph{incremental potential} (IP).
The idea was later picked up in computer graphics \cite{KYT+06, MTG+11, LBO+13}, and the resulting \emph{optimization-based time integrators}
enabled off-the-shelf as well as custom smooth optimization algorithms to be used, facilitating greater robustness and trade-offs between speed and accuracy,
for example in the form of improved Newton solvers \cite{GSS15}, first-order methods~\cite{BML+14, OBL+17, MEM+20, WY16}, quasi-Newton methods~\cite{LBK17, LGL+19} and specialized methods~\cite{LLJ+23}.
The incremental potential concept has since been applied to a great variety of simulation problems, including but not limited to elastic solids with and without contact \cite{LFS+20, LMY+22}, cloth and shells~\cite{LKJ21, CXY+23}, rods
~\cite{LKJ21}, rigid bodies~\cite{LKL+22, FLS+21, CLL+22}, snow~\cite{GSS15} and fluids~\cite{XLY+23, RGJ+15}.

We focus exclusively on Newton-type solvers that incorporate a line search applied to the incremental potential.
Our evaluation is focused on volumetric deformable solids discretized by the Finite Element Method (FEM), but we do not believe that our findings are intrinsic to this context; we expect many of our insights to carry over to other physical systems.

\paragraph{Semidefinite projection}
The incremental potential $E$ is often non-convex, and so the Hessian $\vec H$ may be indefinite.
Motivated by the desire to use the Conjugate Gradient method~\cite{HS52}, which requires the system matrix to be positive definite, \citet{TSI+05} proposed to modify the matrix assembly procedure for the Hessian matrix used in Newton's method so that each \emph{element matrix} associated with the elastic potential for a finite element model is projected to positive semidefiniteness, i.e. its eigenvalues are clamped to be non-negative.
For dynamic problems, this is sufficient for the global matrix to be positive definite.
Later, researchers extended and improved upon this \emph{analytic} eigendecomposition procedure for isotropic~\cite{MZS+11, SHS+12, XSZ+15, SGK18, SGK19, LCK22} and anisotropic~\cite{KGI10} materials, cloth/shells~\cite{Kim20, Pan20} and collision energies~\cite{SK23}.

\paragraph{Projected Newton}
In addition to allowing the use of the Conjugate Gradient method, a positive definite system matrix also ensures that the modified Newton search direction points in a direction of energy decrease.
Coupled with a line search, the resulting modified Newton method has in recent years increasingly been referred to as \emph{Projected Newton} (PN), possibly first described this way by \citet{SPS+17}, and subsequently in a number of papers on geometry processing~\cite{MKK+18, SBC+19, DAZ+20} and simulation~\cite{WLF+20, LSD+22, Dav20} (not an exhaustive list).
Unfortunately, this is the same name given to a widely used class of optimization methods dating back to the early 80s~\cite{Ber82}.
To avoid this clash, we suggest to use a different name, such as \emph{Filtered Newton}.
Nevertheless, we will continue to use the term \emph{Projected Newton} in this paper.

To the best of our knowledge, all works that use Projected Newton apply semidefinite projection \emph{unconditionally} at each iteration.
Contrary to what may be considered accepted belief, our experiments indicate that this strategy may often perform considerably worse than Newton's method.

We remark that we solve the incremental potential to a certain precision; this observation does not necessarily apply to applications that use a fixed number of iterations of Projected Newton (perhaps just one), and/or use different metrics for what constitutes an acceptable solution (such as the visual outcome).
For example, Pixar's Fitz simulation system for visual effects uses a single PN iteration at each step~\cite{KE22}.

\paragraph{Alternatives to PN}
In the context of Newton-type methods, there are --- broadly speaking --- two classes of methods: line search methods and trust region methods \cite{NW06}.
Trust region methods are considerably more complex for non-convex problems, and to our knowledge have not seen much use for simulation in graphics.
A further benefit of line search methods is the ability to incorporate \emph{Continuous Collision Detection} to prevent tunneling in simulations with contact~\cite{LFS+20}.
We therefore focus on line search methods, and the difference between alternatives then essentially boils down to the Hessian approximation at each iteration.

A simple alternative to semidefinite projection is to regularize the Hessian by adding a multiple $\sigma$ of a diagonal matrix $\vec D$, such that $\vec H + \sigma \vec D$ becomes positive definite~\cite{NW06, MTG+11}.
However, it is difficult to set an \emph{initial} value for $\sigma$, and it is not clear how to select $\vec D$ for the simulation problems we are interested in.
We show in \sectionref{sec:kinetic_newton} how we can exploit the structure of the incremental potential problem to eliminate the need for an initial $\sigma$, leading eventually to our Kinetic Newton method.

Another view of this idea is that of \emph{regularization}.
The idea of regularizing Newton's method dates back to the celebrated work of Levenberg and Marquardt on least-squares problems~\cite{Lev44, Mar63},
forming the foundation for the \emph{trust-region} method applied to general  unconstrained optimization problems~\cite{CGT00}.
However, the Levenberg-Marquardt (LM) regularization strategy has proven useful outside the trust-region framework, inspiring new directions of research in Newton-type methods \cite{NP06, Pol09, Mis23}.
Our Kinetic Newton algorithm is closely related to LM regularization, but specialized to the structure of incremental potentials.

Newton will often produce descent directions even when the Hessian is indefinite.
\citet{GSS15} suggest to \emph{flip} the search direction for Newton's method if it is an ascent direction.
This does not ensure a \emph{good} step, but it makes it possible to use the exact Hessian in a line search solver even when indefinite, and we use the same technique for our baseline Newton solver.
If the (possibly flipped) Newton direction is almost orthogonal to the negative gradient, they revert to the steepest descent direction.
However, such angle tests may reject good steps and drastically reduce convergence for ill-conditioned problems, and so are problematic for challenging problems~\cite[Section 3.3]{NW06}.
We do not incorporate an angle test in our baseline Newton method.

\paragraph{Inexact Newton methods}
It is common to use iterative linear solvers to only \emph{approximately} compute the Newton step~\cite[Chapter 7]{NW06}.
Parameters like the choice of iterative solver, stopping tolerance, maximum number of iterations and preconditioner may all influence the convergence of the Newton solver, which severely complicates analysis.
Therefore, we exclusively consider direct solvers for solving the linear systems that arise, so as to get as close as possible to the behavior of \emph{exact} Newton methods.

\paragraph{Convergence criteria, line search}
Related works for our robust line search (\sectionref{sec:robust_line_search}) and our discussion on convergence criteria (\sectionref{sec:convergence_criteria}) are discussed directly in the respective sections.
\section{Fundamentals}

In this section, we review the concept of \emph{incremental potentials}, a class of optimization problems that arise from the reformulation of the discrete equations of motion for a particular time integrator, such as Backward Euler.
Next, we summarize relevant basic optimization theory for Newton's method (see e.g.~\citet{NW06}).

\subsection{Incremental Potential}
We consider discrete time-stepping problems described by an Ordinary Differential Equation (ODE) of the form
\begin{align}
\label{eq:ode}
\begin{aligned}
\vec M \dot{\vec v}(t) &= \vec f(t, \vec u, \vec v) := -\nabla_{\vec u} \phi(t, \vec u) - \nabla_{\vec v} W(t, \vec v)
\\
\qquad \dot{\vec u}(t) &= \vec v(t).
\end{aligned}
\end{align}
Here $\vec M$ is the positive definite mass matrix and the state vector $\vec u$ is a vector containing the positional degrees of freedom for the problem, and $\vec v$ a vector containing the velocity degrees of freedom.
For example, $\vec u$ could contain the nodal displacements for a deformable solid discretized in space by the Finite Element Method (FEM), or an aggregation of degrees of freedom from bodies in a multibody system.

The assumptions we make on the form of $\vec f(t, \vec u, \vec v)$ indicate that it is a separable function in $\vec u$ and $\vec v$.
We can obtain the force from the negative derivatives of a known potential $\phi(t, \vec u)$ and a \emph{dissipation function} $W(t, \vec v)$ (see e.g. \cite{BOF+18, SO18}).
For example, $\phi$ might represent the strain energy of an elastic solid, and $W$ might represent certain forms of damping.

These assumptions do not hold for general dissipative forces like damping and friction, as they are not separable in $\vec u$ and $\vec v$.
For example, even simple Rayleigh damping in the form $\vec D(\vec u) \vec v$ depends both on $\vec u$ and $\vec v$, and so does not satisfy the assumption.
A common workaround is to fix $\vec u$ at the initial state $\vec u^n$ of the current time step, so that we can define $W(t, \vec v) = \vec D(\vec u^n) \vec v$.
Similarly, the \emph{lagged friction} approach taken by IPC~\cite{LFS+20} can also be represented in the form $W(\vec v; \vec u^n)$, though representing friction this way does have caveats~\cite{LLA+22}.

Although this formulation is general and encompasses many types of multiphysics systems, we restrict our evaluation to single- and multibody simulation of deformable solids.

\subsubsection{Backward Euler} We now abuse notation to use $\vec u$ in place of $\vec u^{n + 1}$ as we apply the standard Backward Euler integrator to \eqref{eq:ode} and obtain the nonlinear system
\begin{align}
\label{eq:backward_euler_residual}
\vec r(\vec u)
= \vec M \frac{\vec u - \tilde{\vec u}}{(\Delta t)^2} - \vec f \left(t^{n+1}, \vec u, \frac{\vec u - \vec u^n}{\Delta t} \right)
= 0,
\end{align}
where $\vec r = \vec r(\vec u)$ denotes the \emph{residual}, and $\tilde{\vec u} = \vec u^n + \Delta t \vec v(t^n)$
\footnote{
Constant external forces are often included in $\tilde {\vec u}$ as $\vec M^{-1} \vec f_\text{ext}$, but this may require a mass matrix inversion, which may be impractical if mass lumping is not or cannot be used. Instead, we assume that constant forces are incorporated into $\phi$.}.
Once a solution $\vec u^{n + 1}$ has been found, the velocity is updated according to the formula
\begin{align}
\vec v^{n + 1} = \frac{\vec u^{n + 1} - \vec u^n}{\Delta t}.
\end{align}
The residual equation \eqref{eq:backward_euler_residual} is a general nonlinear problem and can be solved with a root finding algorithm. An advantage of time-stepping problems is that the solution at one time step tends to lie close to that of the previous time step (\emph{temporal coherence}), and so root finding algorithms such as appropriately damped variants of Newton's method will often converge. However, this cannot be guaranteed, and for challenging problems, directly solving for a root may be computationally intractable.

Instead, we take advantage of our assumptions on $\vec f(t, \vec u, \vec v)$ to reformulate the Backward Euler formula \eqref{eq:backward_euler_residual} as
\begin{align}
\label{eq:backward_euler_optimality}
\vec r(\vec u) =
\vec M \frac{\vec u - \tilde{\vec u}}{(\Delta t)^2}
+ \nabla_{\vec u} \phi \left( t^{n+1}, \vec u \right)
+ \nabla_{\vec v} W \left( t^{n + 1}, \frac{\vec u - \vec u^n}{\Delta t} \right)
= 0,
\end{align}
which happens to be the first-order optimality condition for the energy minimization problem
\begin{align}
\label{eq:backward_euler_incremental_potential}
\min_{\vec u} E(\vec u)
\end{align}
with
\begin{align}
\begin{aligned}
E(\vec u)
&:= \frac{(\vec u - \tilde{\vec u})^T \vec M (\vec u - \tilde{\vec u})}{2 (\Delta t)^2}
+ \phi(t^{n+1}, \vec u)
+ \Delta t W \left(t^{n+1}, \frac{\vec u - \vec u^n}{\Delta t} \right)
\\
&= E_M(\vec u) + E_\phi(\vec u) + E_W(\vec u).
\end{aligned}
\end{align}
This is the \emph{incremental potential} (IP) for the Backward Euler integrator.
Any stationary point of the IP is a root of the original residual equation \eqref{eq:backward_euler_optimality}.
There exist methods that enjoy convergence to a stationary point from any starting point, provided $E$ is $C^1$ and satisfies mild regularity assumptions (see for example~\cite[Section 3.2]{NW06}).
This is not the case if we try to directly find a root of \eqref{eq:backward_euler_residual} without knowledge of $E$.
In other words, optimizing $E$ instead of finding roots of $\vec r$ enables us to formulate methods that have much stronger convergence properties.
Moreover, if $\phi$ and $W$ are only $C^1$ smooth, then minimizing $E$ still falls within the realm of smooth optimization, whereas $\vec r$ is $C^0$ and therefore non-smooth.
For non-smooth equations, the local convergence guarantee associated with line search methods goes away, and so root finding is even more likely to slow down or stagnate.

\sloppy Other integrators may admit similar incremental potentials~\cite{BOF+18, DLK18}, however we restrict our attention to Backward Euler to limit the scope of the experimental evaluation.

\subsection{Newton's method}
In the following, we will let $n$ represent a time step counter and $k$ an iteration counter, so that $\vec u^k$ denotes the current estimate of $\vec u^{n+1}$ for the $k$th iteration of Newton's method. The step $\Delta \vec u$ seeks to compute the minimum of the quadratic approximation
\begin{align}
\label{eq:newton_subproblem}
\min_{\Delta \vec u}
\nabla E \cdot \Delta \vec u
+ \frac{1}{2} \Delta \vec u^T \vec H \Delta \vec u,
\end{align}
where $\nabla E = \nabla E(\vec u^k)$ and $\vec H = \vec H(\vec u^k)$ is the Hessian of $E$. The Newton step $\Delta \vec u$ is given by
\begin{align}
\Delta \vec u = - \vec H^{-1} \nabla E(\vec u^k),
\end{align}
and if $\vec H$ is invertible, it is a stationary point of the quadratic problem.
The current solution estimate is then updated to $\vec u^{k + 1} := \vec u^k + \Delta \vec u$.

Even if $E$ is strongly convex, Newton's method may diverge.
To ensure convergence, we incorporate a backtracking line search.
The line search finds a parameter $\alpha$ such that $\vec u^{k + 1} \gets \vec u^k + \alpha \Delta \vec u$ achieves a \emph{sufficient reduction} at each iteration.

If $E$ is non-convex, $\vec H$ may be indefinite, and the Newton step may not be a descent direction.
In this case, the line search may not find \emph{any} acceptable $\alpha$ and the optimization process stalls.
Following \citet{GSS15}, we flip the direction $\Delta \vec u$ if it is an ascent direction, which gives a descent direction $-\Delta \vec u$.
This does not guarantee that the step is \emph{good}, and in theory it may stall altogether if the initial step is neither descent nor ascent, i.e. $\nabla E \cdot \Delta \vec u = 0$.
However, we observe that for incremental potentials it almost always allows the optimization process to continue until Newton gets close enough to converge faster.
This lets us use Newton's method as a baseline for comparison, even when it produces an ascent direction.

\subsubsection{Affine invariance and asymptotic mesh independence}
\label{sec:newton_affine_invariance}
Challenging problems, especially those involving stiff materials and large deformation, tend to exhibit ill-conditioned Hessian matrices.
Another source of ill-conditioning is \emph{resolution}.
The discretization of elliptic-like PDEs with the Finite Element Method produces Hessians that tend to get increasingly ill-conditioned as the resolution improves.

First-order methods, such as gradient descent, generally suffer reduced convergence rates as the Hessian becomes more ill-conditioned~\cite{NW06, Alg19}. While each iteration of Newton's method is considerably more expensive, it is largely invariant to ill-conditioning, which makes it a good candidate for difficult problems. The explanation for this behavior is \emph{affine invariance}, in short the property that Newton's method behaves essentially the same under an affine transformation of coordinates~\cite{Deuflhard2005}.

Newton's affine invariance further gives rise to a property often referred to as \emph{asymptotic mesh independence}~\cite[Chapter 8.1]{Deuflhard2005}.
Roughly speaking, if the discrete nonlinear system stems from a convergent discretization of a PDE, and if the nonlinearity is sufficiently mild, then the convergence behavior will tend to be similar beyond a certain resolution.
\section{Projected Newton}
In this section, we first review the Projected Newton method. Then we empirically study the \emph{negative} effects the semidefinite projection may have on convergence through the lens of the simple \emph{swinging beam} example mentioned in the introduction and presented in \figref{fig:teaser} (Right). To our knowledge, these drawbacks of Projected Newton relative to Newton's method have not previously been documented in the literature.

\label{sec:projected_newton}

\subsection{Background}
\label{sec:projected_newton_background}

Newton's method may stall or slow down when $\vec H$ is not sufficiently positive definite (all eigenvalues $\lambda_i \gg 0$). Projected Newton addresses this problem by exploiting the structure of the Hessian matrix. We have that
\begin{align}
\label{eq:backward_euler_hessian}
\begin{aligned}
\vec H
= \hessian{E}{\vec u}
&= \hessian{E_M}{\vec u} + \hessian{E_\phi}{\vec u} + \hessian{E_W}{\vec u}
\\
&= \frac{1}{(\Delta t)^2} \vec M + \vec H_\phi + \vec H_W.
\end{aligned}
\end{align}
Since $\vec M$ is positive definite, we can obtain a positive definite approximation of $\vec H$ if we can replace the Hessian terms related to forces, $\vec H_f = \vec H_\phi + \vec H_W$, by a similar positive \emph{semidefinite} matrix.
How exactly to do this is somewhat problem-specific, but the general idea is to rely on the property that, for most problems encountered in multibody simulation, $\vec H_f$ can be written as a sum of low-rank matrices:
\begin{align}
\label{eq:H_f_sum}
\vec H_f = \sum_i \vec R_i^T \vec H_{f,i} \vec R_i.
\end{align}
This is the case whenever individual energy contributions (such as for a single \emph{element}) are only influenced by a small number of degrees of freedom.
Then $\vec H_{f,i}$ is a \emph{small} matrix (for example $12 \times 12$ for the strain energy of a tetrahedral finite element) and $\vec R_i$ is an operator that selects the relevant degrees of freedom for the $i$-th contribution.
We now observe that $\vec H_f$ is semidefinite if all $\vec H_{f,i}$ are.
To ensure that this is the case, each $\vec H_{f,i}$ is \emph{projected} to semidefiniteness in some way, either through a costly numerical eigendecomposition for the general case, or by specialized analytic methods for particular cases such as strain energies for deformable solids~\cite{KE22}.

The result is that the Hessian approximation $\vec H_\text{proj}$ employed by Projected Newton is always positive definite, which together with a backtracking line search guarantees progress towards a minimum at every iteration.

However, quite often, the Hessian matrix $\vec H$ for Newton's method is \emph{already} sufficiently positive definite, even though some of the individual element matrices $\vec H_{f,i}$ are indefinite.
This happens when the indefiniteness is counteracted by the influence of other energies acting on the same degrees of freedom.
Often, the Hessian matrix for the first iteration is the same or very similar to the Hessian for the solution of the previous incremental potential.
If a strict minimum was found, then we will start the new solve with a positive definite matrix.
Moreover, if the time step $\Delta t$ is small enough, then the positive definite mass matrix $\vec M$ has a regularizing influence on the spectrum of $\vec H$.
Unless \emph{all} element matrices $\vec H_{f,i}$ are already positive semidefinite, then $\vec H_\text{proj} \neq \vec H$, even in the vicinity of a minimum.
In general, therefore, we can not expect Projected Newton to possess the affine invariance property of Newton's method, nor inherit the same convergence rate near a minimum.

\subsection{Effect on convergence}
\label{sec:projected_newton_convergence}
To study the effect of the semidefinite projection on convergence, we now revisit and study the \emph{swinging beam} example first presented in the introduction and in
\figref{fig:teaser} (Right) in more detail.
The ${2 \times 1 \times \SI{1}{\meter}}$ beam is discretized in space by regular tetrahedral meshes with various resolutions, and in time with a time step size $\Delta t = \SI{16.7}{\milli \second}$.
We use a Neo-Hookean material model with Young's modulus $E = \SI{0.40}{\mega\pascal}$, Poisson's ratio $\nu = 0.40$ and density $\rho = \SI{1000}{\kilo\gram\per\cubic \meter}$, reminiscent of a rubber-like material.
The beam is clamped on one of its short ends and falls freely under gravity for 6 seconds (see the supplemental video and
snapshots in \figref{fig:swinging_beam_iterations}).
We consider two different values for convergence tolerance, a \emph{coarse} tolerance $\epsilon_a = 1.00$ and a \emph{fine} tolerance $\epsilon_a = 0.01$, where $\epsilon_a$ refers to the acceleration-based criterion that we propose in \sectionref{sec:acceleration_balance}.
See \sectionref{sec:implementation} for additional implementation notes.

\subsubsection{Mesh dependence}

From \figref{fig:teaser} (Right), we observe that the iteration count increases significantly with increased mesh resolution for Projected Newton, but the number of iterations for Newton's method stays almost constant.
This is consistent with the observation that Newton's method enjoys affine invariance (\sectionref{sec:newton_affine_invariance}), but Projected Newton does not (\sectionref{sec:projected_newton_background}).
The results indicate a much stronger dependence on mesh resolution for Projected Newton than for Newton's method, which could degrade its performance for large problems.

\subsubsection{Convergence rate}
\begin{figure}[htb!]
    \begin{center}
        \includegraphics{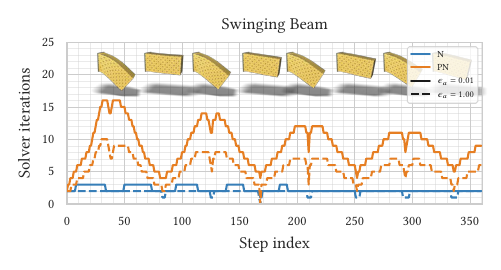}
    \end{center}
    \caption{Iterations for each time step of the swinging beam experiment with 58.3k vertices and 332k elements.
	The deformation of the beam is shown for each peak and valley in the PN iteration plot.
    }
    \label{fig:swinging_beam_iterations}
\end{figure}
~
\figref{fig:teaser} (Right) also indicates that Projected Newton suffers much more from a stricter convergence criterion than Newton's method.
\figref{fig:swinging_beam_iterations} shows that a single additional step is often sufficient for Newton to satisfy a lower error tolerance tolerance ($\epsilon_a = 0.01$) compared to a larger one ($e_a = 1.00$).
The difference is more significant for Projected Newton, suggesting a markedly slower convergence rate.

\figref{fig:swinging_beam_iterations} also shows that the iteration count for Newton is fairly constant throughout the entire experiment, but Projected Newton clearly requires more iterations when the deformation of the beam is larger.
Neo-Hookean materials tend to exhibit strongly negative eigenvalues under large deformation, and so Projected Newton discards a larger part of the eigenspectrum as the deformation grows.
The discarded eigenvalues are likely key to Newton's fast convergence in this experiment.

\subsubsection{Projection method}
\begin{figure}[t!]
    \begin{center}
        \includegraphics{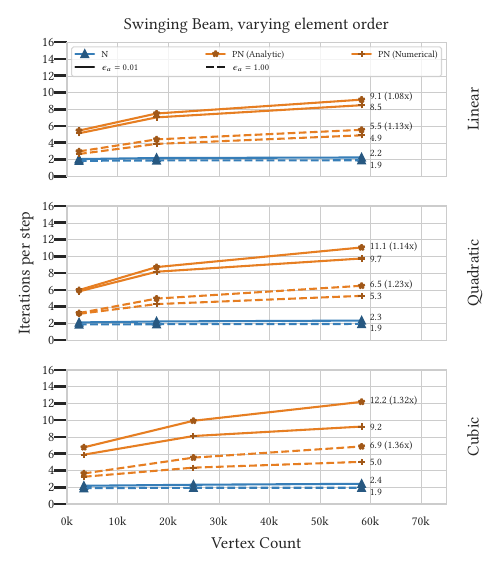}
    \end{center}
    \caption{Average number of solver iterations per time step for linear, quadratic and cubic tetrahedral meshes of varying resolution. For PN, the results for both numerical and analytic projection are presented. The multipliers in parantheses denote the ratio of iterations between analytic and numerical projection for the same configuration.
    }
    \label{fig:swinging_beam_element_order}
\end{figure}
There are sometimes several ways of performing the semidefinite projection.
~
\begin{figure}[t!]
    \begin{center}
        \includegraphics{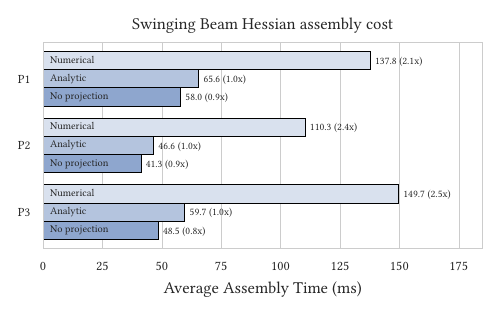}
    \end{center}
    \caption{Average cost of global Hessian assembly for the highest resolutions of the Swinging Beam experiments, for linear (P1), quadratic (P2) and cubic (P3) elements.
    Each bar corresponds to a different kind of semidefiniteness projection method.}
    \label{fig:swinging_beam_assembly_cost}
\end{figure}
~
For example, for deformable FEM problems, the matrix $\vec H_{f, i}$ from \eqref{eq:H_f_sum} is also composed of a sum of matrices with a particular structure.
In such cases $\vec H_{f,i}$ can be decomposed as the sum of $Q_i$ contributions, one for each quadrature point:
\begin{align}
\label{eq:H_f_i_decomposition}
\vec H_{f,i} = \sum_{q=1}^{Q_i} \vec B_{i,q}^T \vec H_{f, i, q} \vec B_{i,q}.
\end{align}
In this case $\vec B_{i,q}$ are rectangular matrices that encode the third order tensor $\pd{\vec F}{\vec u}$. For 3D elastic energies, $\vec H_{f, i, q}$ is a $9 \times 9$ matrix that depends only on material properties, independent of the FEM discretization.
This structure leaves us with two main alternatives for projection:
\begin{itemize}
\item \emph{Numerical.} Assemble $\vec H_{f,i}$ from \eqref{eq:H_f_i_decomposition}, compute a numerical eigendecomposition, project eigenvalues to $[0, \infty)$ and rebuild the matrix.
\item \emph{Analytic.} Use an \emph{analytic} eigendecomposition of $\vec H_{f,i,q}$ to project $\vec H_{f,i,q}$ to semidefiniteness, which implies semidefiniteness of $\vec H_{f,i}$.
\end{itemize}
For background, we recommend the course by \citet{KE22}, which offers a detailed treatment of both the matrix structure and state-of-the-art techniques for analytic projection.

We will study the impact this choice has on convergence in the context of tetrahedral finite elements of different polynomial order, where $Q_i$ increases with element order.
\figref{fig:swinging_beam_element_order} demonstrates the average number of iterations per time step for different resolutions and projection methods, for linear, quadratic and cubic elements.
From the results, there is a clear trend for analytic projections to require increasingly more iterations relative to numerical projections as the element order (and therefore $Q_i$) increases, requiring as much as $36\%$ more iterations compared to numerical projection for cubic elements.

The difference between numerical and analytic eigenprojection is analogous to the difference between $\max(0, \sum_i x_i)$ and $\sum_i \max(0, x_i)$ for finding a non-negative approximation of a sum $\sum_i x_i$.
By that analogy, the numerical projection is generally closer to the original sum, which likely explains the discrepancy.
In other words, analytic projection will remove more of the negative eigenvalue information that we have already found to be essential for fast convergence.

We present the average runtime cost of global Hessian assembly for each element type in \figref{fig:swinging_beam_assembly_cost}, where we have only considered the highest resolution meshes that all have 58.3k vertices.
The element counts are 332k (linear), 41.5k (quadratic) and 12.3k (cubic).

The numbers only consider the cost of computing the Hessian of the elastic strain energy in parallel, using a pre-computed graph coloring and sparsity pattern.
For linear elements, we note a distinct gap in runtime between analytic and numerical projection (2.1x), but despite the algorithmic complexity of numerical eigendecomposition, the relative gap only slightly increases with higher order.

The cost of assembly relative to other parts of a simulation is strongly application-dependent, and depends on many factors. The results here suggest that numerical projection may sometimes be favorable when taking into account the reduced iteration counts, which to our knowledge is a novel insight.

\subsubsection{Material model}
We repeat the experiments with linear elements for the Stable Neo-Hookean material model~\cite{SGK18}, which is a more numerically well-behaved model than Neo-Hookean.
The iteration counts are almost identical to the results of the experiments with the Neo-Hookean model (see \appendixref{sec:swinging_beam_stable_neo_hookean}),
confirming that the observed behavior is not a quirk of the Neo-Hookean model.
\section{Projection on demand}
\label{sec:projection_on_demand}
Our investigation in \sectionref{sec:projected_newton_convergence} revealed that semidefinite projection sacrifices important characteristics of Newton's method.
At the same time, Newton's method can be unreliable in settings where the Hessian might become significantly indefinite.
In this section, we will propose a simple method that seeks to combine the best of both.

Ideally, we would like to take unmodified Newton steps whenever Newton steps are \emph{good}, and otherwise apply projection to help us move towards a region of positive definiteness, so that Newton can get us the rest of the way.
This line of thinking leads to what \citet[Chapter 3.4]{NW06} call \emph{Newton's method with Hessian modification}.
The gist is that we replace the exact Hessian $\vec H(\vec u^k)$ at iteration $k$ of Newton's method with a matrix $\vec B_k = \vec H(\vec u^k) + \vec E_k$, where $\vec E_k$ is chosen such that $\vec B_k$ is sufficiently positive definite.
In particular, if $\vec H(\vec u^k)$ \emph{is} already acceptably positive definite, we set $\vec E_k = 0$ and recover Newton's method in convex regions.

The simplest modification of Projected Newton necessary to fit such a scheme would be to try to compute a sparse Cholesky factorization of $\vec H$.
If it succeeds, then take a full Newton step ($\vec B_k = \vec H(\vec u^k)$), otherwise compute $\vec H_\text{proj}$ and compute the corresponding step ($\vec B_k = \vec H_\text{proj}(\vec u^k)$).

There are two issues with this approach.
First, if we are currently moving through a non-convex region (i.e. $\vec H$ is indefinite), then at each iteration we compute $\vec H$ and quickly discard it in favor of $\vec H_\text{proj}$, which must be computed from scratch.
This is clearly computationally wasteful.
Second, it might be that the Cholesky factorization succeeds, but $\vec H$ is still close to singular and the resulting Newton step may be poor.

The latter situation is not easy to detect reliably in a computationally tractable manner.
As a cheap heuristic, we have found that monitoring $\alpha$, the step length factor returned by the line search, seems to work well.
The idea is that a poor step \emph{likely} returns $\alpha < 1$, and we use this as an indicator that we should apply semidefinite projection in the next step.
This also means that we will use projection if the full step is rejected due to CCD.
It would also be possible to consider the ratio $\alpha_\text{backtracking} / \alpha_\text{CCD}$, but this would slightly complicate the method since CCD cannot be incorporated \emph{inside} the line search, and at least in the context of IPC contacts, $\alpha_\text{CCD} < 1$ anyway often indicates that the incremental potential $E$ is strongly non-convex in this region.

While monitoring $\alpha$ helps prevent excessive numbers of failed Newton steps, it is quite typical that a projected Newton step will return $\alpha = 1$, only for the subsequent attempt with a full Newton step to fail due to indefiniteness of $\vec H$.
This will typically lead to many unnecessary matrix assemblies and solver attempts.
To remedy this, we introduce the rule that, whenever an indefinite matrix is encountered, we take at least $N$ projected steps before trying a full Newton step again.
We find that $N = 4$ seems to strike a good balance; it lets us amortize the cost of failed assembly and factorization over several steps, while recovering full Newton steps fairly quickly if we manage to get close enough for Newton's method to be effective.

This algorithm, which we refer to as \emph{Project-on-Demand Newton} (POD-Newton), is summarized in \algoref{alg:project_on_demand_newton}.

\begin{algorithm}[htb]
\caption{The Project-on-Demand Newton method. $\vec H^{-1}$ denotes a linear operator represented implicitly by a factorization.}
\begin{algorithmic}
\State \texttt{project\_psd} $\gets$ \texttt{false}
\Comment{Whether to project at next iteration}
\State \texttt{countdown} $\gets 0$
\While{not converged}
	\State Assemble $\vec H$
	\Comment {Project if \texttt{project\_psd} is \texttt{true}}
	\If{$\vec H^{-1} \gets \texttt{factor}(\vec H)$ fails}
	\Comment{e.g. Cholesky}
		\State \texttt{countdown} $\gets 3$
		\Comment{$N - 1$}
		\State \texttt{project\_psd} $\gets$ \texttt{true}
		\State Re-assemble $\vec H$ with PSD projection
		\State $\vec H^{-1} = \texttt{factor}(\vec H)$
	\EndIf
	\\
	\State $\vec \Delta \vec u \gets - \vec H^{-1} \nabla E(\vec u^k)$
	\State $\alpha \gets \texttt{line\_search}(\Delta \vec u)$
	\State $\vec u^{k + 1} \gets \vec u^k + \alpha \Delta \vec u$
	\\
	\State \texttt{project\_psd} $\gets \alpha < 1 \textbf{ or } \texttt{countdown} > 0$
	\State $\texttt{countdown} \gets \max(0, \texttt{countdown} - 1)$
\EndWhile
\end{algorithmic}
\label{alg:project_on_demand_newton}
\end{algorithm}
\section{Kinetic Newton}
\label{sec:kinetic_newton}

As we will see in \sectionref{sec:evaluation}, POD-Newton exhibits fast, reliable convergence for most problems.
However, like Projected Newton, it requires intricate modifications of the Hessian assembly, which may be costly or difficult to implement efficiently.
It may also prevent the use of black-box energies and their derivatives, for example obtained through automatic differentiation.
In fact, POD-Newton adds an additional layer of complexity on top of Projected Newton:
the programming interface between the optimizer and the energy contributions ($\phi$, $W$) must allow for the optimizer to communicate the desired definiteness of the Hessian at each call to the assembly routine.

It would be convenient if we did not need to concern ourselves with semidefinite projection at all, and we could instead work directly with the actual Hessian $\vec H$ at all times.

We now propose \emph{Kinetic Newton}, a novel regularization-based alternative.
Following Levenberg-Marquardt regularization, we augment the subproblem for the Newton step \eqref{eq:newton_subproblem} with a regularization term, leading to a regularized subproblem for the $k$th step $\Delta \vec u$:
\begin{align}
\label{eq:newton_regularized_subproblem}
\min_{\Delta \vec u}
\nabla E \cdot \Delta \vec u
+ \frac{1}{2} \Delta \vec u^T \vec H \Delta \vec u
+ \frac{\gamma_k}{2} \| \Delta \vec u \|^2
\end{align}
for some regularization parameter $\gamma_k \geq 0$ and unspecified norm $\| \cdot \|$.
For $\gamma_k = 0$, no regularization is applied, and we recover the standard Newton step. The role of the regularization term is two-fold.
First, the penalty on step length $\| \Delta \vec u \|$ encourages shorter steps.
Second, the regularization term adds a positive definite matrix to the Hessian, which ensures positive definiteness for sufficiently large $\gamma_k$ (\appendixref{sec:pos_def_regularized_Hessian}).

One of the main difficulties of a regularized Newton's method is picking appropriate values of $\gamma_k$.
Ideally we would like to pick $\gamma_k$ large enough to make the modified Hessian sufficiently positive definite, yet not so large as to cause us to take unnecessarily small --- and possibly low quality --- steps.
We can iteratively try to determine a reasonable choice of $\gamma_k$.
If we regularize with the Euclidean norm, this corresponds to adding a multiple of the identity to the Hessian, i.e. the Hessian is replaced by $\vec B_k = \vec H(\vec u^k) + \gamma_k \vec I$.
In order to find $\gamma_k$, a sequence of trial values is attempted until $\vec B_k$ can be factorized with sparse Cholesky, which ensures that $\vec B_k$ is in fact positive definite.

The difficulty with this strategy lies in choosing the initial trial value for $\gamma_k$ and the subsequent strategy for incrementally updating it. If the initial value is too small, many failed factorization attempts might be made.
If the value is too large, it may lead to poor convergence.

We first propose to use the \emph{kinetic norm} induced by the mass matrix,  defined by $\| \Delta \vec u \|^2_M = \Delta \vec u^T \vec M \Delta \vec u$ for the regularization term.
This ensures that the regularization takes into account physical properties of the discretization, so that for example small elements (with smaller mass) are weighted less than large elements (with larger mass).
Next, we set
\begin{align}
\gamma_k = \frac{1 - \beta_k^2}{(\beta_k \Delta t)^2}
\end{align}
for a parameter $\beta_k \in (0, 1]$ that is adjusted adaptively. With this choice, the \emph{regularized Hessian} $\vec H_\beta$, i.e. the Hessian corresponding to the objective function of the regularized subproblem \eqref{eq:newton_regularized_subproblem}, is
\begin{align}
\vec H_\beta
= \frac{\vec M}{(\Delta t)^2} + \vec H_f + \gamma \vec M
= \frac{\vec M}{(\beta \Delta t)^2} + \vec H_f.
\end{align}
In the absence of a dissipation function, i.e. $W = 0$, then $\vec H_\beta$ exactly coincides with the Hessian of the Newton subproblem \eqref{eq:newton_subproblem} for a \emph{smaller} time step $\beta \Delta t$.

Reducing the size of the time step is a tried and tested technique for solving difficult time stepping problems.
Smaller time steps both improve the conditioning of the Hessian (since the mass matrix is weighted more heavily) and reduce nonlinearity at each time step.
The connection we have established to our regularization scheme suggests that we might use a similar mechanism for regularizing the Hessian, but without actually restarting the solution process with a smaller time step.
Instead, we reduce or increase $\beta_k$ as necessary to ensure continued progress.
Since we only change the Hessian and not the gradient, we are still solving the same problem but with a different Hessian approximation so that the steps may be considered as something of a mix between gradient descent preconditioned with $\vec M^{-1}$ ($\beta \rightarrow 0$) and the full Newton step ($\beta = 1$).

We generally start with $\beta_0 = 1$ at the beginning of a time step to indicate no regularization. Whenever factorization of $\vec H_\beta$ fails, we reduce $\beta$ by a constant factor 2.
Similar to POD-Newton, we use the result of line search to decide how to adjust the regularization.
Our proposed method therefore eliminates the need for an initial value for $\gamma$, since we only need to decide on the rate at which to reduce $\beta$.
The method is summarized in \algoref{alg:kinetic_newton}.

The exact values chosen for deciding when to adjust $\beta$ based on $\alpha$ do not appear to matter all that much.
We observed that when $\alpha = 0.5$ is chosen for a full Newton step, then often $\alpha = 1$ follows in the next iteration.
If we add regularization for $\alpha = 0.5$, then we might delay successful Newton steps.
The values used for adjusting $\beta$ in \algoref{alg:kinetic_newton} reflect these observations.

\begin{algorithm}
\caption{The Kinetic Newton algorithm applied to a single time step of Backward Euler at time $t^n$.}
\begin{algorithmic}
\State $\beta \gets 1$
\State Assemble $\vec M$ (or re-use from earlier time step)
\While{not converged}
	\State Assemble $\vec H_f(\vec u^k)$
	\State $\vec H_\beta = \frac{\vec M}{(\beta \Delta t)^2} + \vec H_f$
	\While{$\vec H_\beta$ is indefinite}
		\Comment{e.g. Cholesky failure}
		\State $\beta \gets \beta / 2$
		\State $\vec H_\beta = \frac{\vec M}{(\beta \Delta t)^2} + \vec H_f$
	\EndWhile
	\State $\vec \Delta \vec u \gets - \vec H_\beta^{-1} \nabla E(\vec u^k)$
	\State $\alpha \gets \texttt{line\_search}(\Delta \vec u)$
	\State $\vec u^{k + 1} \gets \vec u^k + \alpha \Delta \vec u$
	\\
	\If{$ \alpha < 0.3 $}
		\State $\beta \gets \beta / 2$
	\ElsIf{$\alpha > 0.9$}
		\State $\beta \gets \min(1, 2 \beta)$
	\EndIf
\EndWhile
\end{algorithmic}
\label{alg:kinetic_newton}
\end{algorithm}

Finally, if $W \neq 0$, exact equivalence no longer holds because $\Delta t$ appears in the expression for $\vec H_W$, however this appears to matter little, as our scheme is anyway only a heuristic for choosing a regularization parameter.

\subsubsection*{Compared to changing the time step}
Given Kinetic Newton's \emph{similarity} to adaptive time-stepping, we outline some differences to the common approach of reducing the time step when the optimizer struggles.
\begin{itemize}
\item In contrast to Kinetic Newton, reducing $\Delta t$ discards current progress.
\item Reducing $\Delta t$ changes the discrete problem.
Not all problems are suited to dynamic timestepping.
\item Reducing $\Delta t$ replaces one optimization problem with several, each of which might need to be solved to high accuracy, compared to coarse intermediate approximations for Kinetic Newton.
\end{itemize}
Smaller time steps may sometimes accelerate the solution process, but it is often (much) more costly.
This depends strongly on the application.
Like \citet{GSS15}, we prefer that the time step be chosen primarily based on the \emph{physical} requirements of the simulation.

\section{Robust line search}
\label{sec:robust_line_search}

If the line search \emph{fails} --- it returns a value of $\alpha$ that is tiny --- it is an indicator that the candidate step $\Delta \vec u$ is bad, and the optimization process stalls unless we change the way in which we compute $\Delta \vec u$.

Unfortunately, in inexact arithmetic, line search may sometimes fail simply because of a numerical deficiency in the Armijo condition, and it can reject valid steps close to a solution.
This artificial failure mode may force us to terminate the simulation without satisfying termination criteria, and makes it harder to analyze convergence failure.

In this section, we discuss the cause of this phenomenon and propose a novel variant of the backtracking line search that essentially removes the artificial failure mode.

\subsection{Numerical instability in the Armijo condition}
\begin{figure}[t!]
    \begin{center}
        \includegraphics{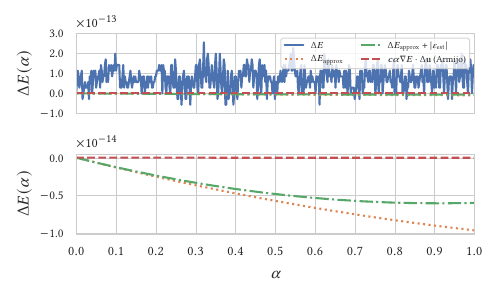}
    \end{center}
    \caption{
    \textbf{Top}: A plot of the value of $\Delta E(\alpha)$ at a point $\vec u$ where the standard backtracking line search fails.
    Numerical errors dominate the standard line search, so that very few values of $\alpha$ satisfy the Armijo condition.
    \textbf{Bottom}: The same plot, but zoomed in and showing only the approximation $\Delta E_\text{approx}$ and our conservative estimation $\Delta E_\text{approx} + |\varepsilon_\text{est}|$.
    Both approximations are smooth and satisfy the Armijo sufficient descent condition, and the conservative estimate converges to $\Delta E_\text{approx}$ as $\alpha$ gets smaller.
    }
    \label{fig:robust_line_search_example}
\end{figure}

The Armijo condition used in backtracking ~\cite[Chapter 3.1]{NW06} accepts a trial step $\Delta \vec u$ with parameter $\alpha$ if it satisfies
\begin{align}
\Delta E(\alpha) := E(\vec u^k + \alpha \Delta \vec u) - E(\vec u^k)
\leq c \, \alpha \nabla E(\vec u^k) \cdot \Delta \vec u
\leq 0
\end{align}
for a fixed constant $c$.
We use $c = 10^{-4}$ throughout.
If $\alpha \Delta \vec u$ is small, then $E(\vec u^k + \alpha \Delta \vec u) \approx E(\vec u^k)$, and inaccuracies in the evaluation may give $\Delta E(\alpha) > 0$ for essentially \emph{any} value of $\alpha$, even though the \emph{exact} value would satisfy the Armijo condition.
Figure \ref{fig:robust_line_search_example} shows a plot of $\Delta E(\alpha)$ at the point where the backtracking line search fails for the \emph{Basic Impact} example with the stiff material (\sectionref{sec:basic_impact}).
The plot shows that the straightforward evaluation of $\Delta E$ is very noisy, and most of the values of $\Delta E(\alpha)$ are \emph{positive} due to numerical error, meaning that the step would be rejected for almost all values of $\alpha$.

This means that candidate solutions that are very close to an exact solution $\vec u^*$ may be rejected simply because $\Delta E$ is too small relative to the precision of $E$.
This is often a hint that the convergence criterion may be overly strict, but this can be difficult to ascertain.
We demonstrate in the \emph{Basic Impact} experiment in \sectionref{sec:evaluation_contact} that for an example with a very stiff material, the standard line search may fail even for a coarse convergence tolerance.

In our experience, the line search will fail due to imprecision long before the gradient gets too inaccurate for the Newton step or the convergence criterion to be useful.
The reason for this is that, as $\vec u \rightarrow \vec u^*$, the converging process $E \rightarrow E^*$ requires more and more digits to be accurate, and so $E$ changes less and less the closer we get to the solution.
In contrast, $\nabla E \rightarrow 0$ means that the gradient will often change drastically from step to step, even when close to the solution.
Thus, the gradient values do not require anywhere near the same degree of precision.

\begin{figure}[htb!]
\definecolor{SmallBoxColor}{RGB}{70,130,180}
\definecolor{LargeBoxColor}{RGB}{244,164,96}

\begin{tikzpicture}

\def\myopacity{0.2}
\def\colorSmallBox{SmallBoxColor}
\def\colorLargeBox{LargeBoxColor}

\def\smallsize{1}
\def\largesize{3}
\def\smalldeform{0.95} %
\def\largedeform{0.5} %
\def\spacing{0.5}
\def\horizontalspacing{\spacing}
\def\eqboxheight{1}

\draw[fill=\colorSmallBox!50, opacity=\myopacity] (0,0) rectangle (\smallsize,\smallsize);
\draw[fill=\colorSmallBox] (0,0) rectangle (\smallsize,\smallsize*\smalldeform) node[pos=0.5] {A};

\draw[fill=\colorLargeBox!50, opacity=\myopacity] (3*\smallsize+\horizontalspacing,0) rectangle (3*\smallsize+\horizontalspacing+\largesize,\largesize);

\coordinate (bottomleft) at (3*\smallsize+\horizontalspacing,0);
\coordinate (bottomright) at (3*\smallsize+\horizontalspacing+\largesize,0);
\coordinate (topright) at (3*\smallsize+\horizontalspacing+\largesize - 0.15,\largesize*\largedeform);
\coordinate (topleft) at (3*\smallsize+\horizontalspacing + 0.15,\largesize*\largedeform);

\coordinate (topcenter) at ($(topleft)!0.5!(topright)$);
\coordinate (leftcenter) at ($(bottomleft)!0.5!(topleft)$);
\coordinate (rightcenter) at ($(bottomright)!0.5!(topright)$);

\def\outbulge{0.5}

\draw[fill=\colorLargeBox]
(bottomleft) .. controls
    +(-\outbulge,0.0*\largesize*\largedeform)
    and +(0,0)
    ..
(topleft) .. controls +(0.5*\largesize,-0.5*\largesize*\largedeform) and +(0, 0) ..
(topright) .. controls +(\outbulge,-1*\largesize*\largedeform) and +(0, 0) ..
(bottomright) --
cycle node {};

\node at (5, 0.5) {$B$};

\draw (0,\spacing + \smallsize) rectangle (3*\smallsize,\largesize) node[pos=.5] {
$\begin{aligned}
E^A &\ll E^B \\
\vec r^A &\neq 0 \quad \vec r^B = 0
\end{aligned}$
};

\end{tikzpicture}
\caption{
A small, soft and barely deformed box (A) and a large, stiff box undergoing large deformation (B).
The large box is at equilibrium (zero residual $\vec r^B = 0$), but its incremental potential energy is very high due to the massive stress.
In contrast, the small box is not in equilibrium, and its energy is very low.
Transparent boxes indicate rest shape.
}
\label{fig:box_example}
\end{figure}
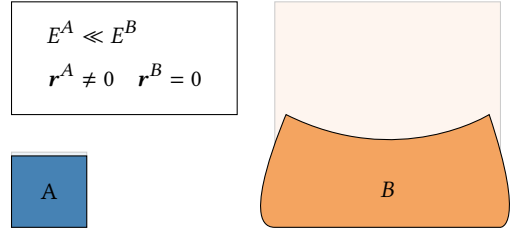

\paragraph{Example}
Consider two solid boxes $A$ and $B$ undergoing various degrees of deformation, as depicted in \figref{fig:box_example}.
The larger box $B$ is in equilibrium ($\vec r^B = \nabla_{\vec u^B} E = 0$), while the smaller box $A$ is not.
Because $B$ is much stiffer, larger and more deformed than $A$, its incremental potential is also much larger, i.e. $E^B \gg E^A$.
Since $\norm{\nabla E} = \norm{\nabla_{\vec u^A} E}$, we have not yet converged, but the floating-point sum $E^A + E^B$ discards many significant digits in the value of $E^A$, because a very small number is added to a very large number.
Eventually, as $\alpha \Delta \vec u$ gets small enough, the change in $E^A(\vec u^A + \alpha \Delta \vec u^A)$ relative to $E_B(\vec u^B)$ resides entirely in the discarded digits, so that $E(\vec u^k + \alpha \Delta \vec u) = E(\vec u^k)$ for any smaller $\alpha > 0$ and we may never satisfy the Armijo condition.

Though contrived, the above example demonstrates how energy contributions of vastly different magnitude can lead to problems when the computation of $E$ is inexact.
The situation described above could be changed to include resting contact between the two solids, or $A$ and $B$ could even be different parts of the same solid, in which all but a single element in the solid has reached equilibrium.
Moreover, energies can be imprecise or reach arbitrarily large values for a variety of other reasons.
For example, constant external forces like gravity are often included as energy terms of the form $-\vec u^T \vec f_\text{ext}$, which can grow very large in magnitude as displacement grows, despite the force remaining constant throughout the simulation.
For this particular case, we recommend to reformulate it as $-(\vec u - \vec u^n)^T \vec f_\text{ext}$, which has the same gradient but only grows roughly proportionally to the delta displacement in a given time step instead of the total displacement.

\subsection{Robust evaluation of the Armijo condition}
We have seen that $\Delta E$ may be numerically inaccurate when $\alpha \Delta \vec u$ is small.
The idea proposed by \citet{HZ05} and later adapted to the trust region setting by \citet{DUS17} is to devise an alternative approximation $\Delta E_\text{approx}$ to $\Delta E$ that is accurate precisely when $\alpha \Delta \vec u$ is small, and is given by elementary Taylor series approximation:
\begin{align}
\label{eq:dE_approx}
\Delta E(\alpha) = \overbrace{\alpha \Delta \vec u \cdot \frac{\nabla E(\vec u^k) + \nabla E(\vec u^k + \alpha \Delta \vec u)}{2}}^{\Delta E_\text{approx}}
+ \mathcal{O}\Big(\norm{\alpha \Delta \vec u}^3\Big).
\end{align}
Since we only need $\Delta E_\text{approx}$ to be accurate to a few digits for the purpose of line search, we do not need the computation of $\nabla E$ to be very accurate, which is in stark contrast to the original formula for $\Delta E$, for which high precision in $E$ is needed.
However, if $\alpha \Delta \vec u$ is sufficiently large, then this approximation is poor, and the optimization process may diverge.
The solution used by \citet{HZ05} is to terminate the line search when either:
\begin{itemize}
\item The original Armijo condition holds.
\item The Armijo condition with $\Delta E_\text{approx}$ in place of $\Delta E$ holds, and $|\Delta E(\alpha)| \leq \epsilon_\text{rel} \, |E(\vec u^k)|$ for some small constant $\epsilon_\text{rel} > 0$.
\end{itemize}
The latter condition on $\Delta E(\alpha)$ is intended to ensure that the approximate criterion is only used when the step is small enough, as it may otherwise cause divergence.

In practice, we found that for challenging problems, we could not find a single parameter $\epsilon_\text{rel}$ that worked across all of our examples, and we still experienced spurious simulation failures.

To remedy this, we propose a simple \emph{parameter-free} algorithm for robust backtracking. The central idea is to approximate an upper bound on the error introduced by the approximate criterion, which is used to prevent the approximate criterion from being used when $\Delta E$ is too large.
Inspired by adaptive error approximation from ODE theory \cite{AP98}, we consider the difference between two estimates of $\Delta E$ with different asymptotic accuracy as a coarse estimate of the error.
To that end, we use Taylor series expansion to define
\begin{align}
\Delta E_\text{coarse}(\alpha) := \alpha \Delta \vec u \cdot \nabla E(\vec u^k + \alpha \Delta \vec u) \approx \Delta E(\alpha),
\end{align}
\sloppy which is an asymptotically coarser approximation of $\Delta E$ than $\Delta E_\text{approx}$.
We use this to obtain a rough estimate of the error $\varepsilon = \Delta E_\text{approx} - \Delta E \approx \Delta E_\text{approx} - \Delta E_\text{coarse} =: \varepsilon_\text{est}$.
This is not a real error bound, but a convenient approximation.
We use this to develop a modified approximate Armijo condition of the form
\begin{align}
\Delta E_\text{approx}(\alpha) + |\varepsilon_\text{est}| \leq c \alpha \nabla E(\vec u^k) \cdot \Delta \vec u.
\end{align}
While non-negative, the error term $|\varepsilon_\text{est}|$ decays quadratically, compared to linearly for $\Delta E_\text{approx}$ ~(\appendixref{sec:robust_line_search_decay_rates}).
Hence, it will reject large steps, but still allow for small steps to be taken.
For sufficiently small steps, the error term essentially vanishes and we recover $\Delta E_\text{approx}$, which we know does not suffer from the numerical deficiency of the original Armijo condition for small steps.

\begin{algorithm}[htb]
\caption{Robust backtracking line search for the incremental potential energy $E$. $\Delta \vec u$ is assumed to be a descent direction.}
\begin{algorithmic}[1]
\Function{robust\_line\_search}{$\vec u$, $\Delta \vec u$; $E$}
	\State $\alpha \gets \texttt{maximum permissible step} \leq 1$
	\Comment{E.g. CCD}
	
	\Loop
		\State $\Delta E \gets E(\vec u + \alpha \Delta \vec u) - E(\vec u)$
		
		\If{$\Delta E \leq c \, \alpha \, \Delta \vec u^T \nabla E(\vec u)$}
			\Comment{Try Armijo first}
			\State %
			\Return $\alpha$
		\ElsIf{$|\Delta E| \leq 0.1 \, |E(\vec u)|$}
			\label{line:robust_line_search_if}
			\State $\Delta E_\text{approx} \gets \frac{\alpha}{2} \Delta \vec u \cdot (\nabla E(\vec u + \alpha \Delta \vec u) + \nabla E(\vec u))$
			\State $|\varepsilon_\text{est}| \gets \frac{\alpha}{2} |\Delta \vec u \cdot (\nabla E(\vec u + \alpha \Delta \vec u) - \nabla E(\vec u) )|$
			\If{$\Delta E_\text{approx} + |\varepsilon_\text{est}| \leq c \alpha \Delta \vec u^T \nabla E(\vec u)$}
			\State
			\Return $\alpha$
			\EndIf
		\EndIf
		
		\State $\alpha \gets \alpha / 2$
	\EndLoop
\EndFunction
\end{algorithmic}
\label{alg:robust_line_search}
\end{algorithm}

We summarize the method in \algoref{alg:robust_line_search}.
The condition on line~\ref{line:robust_line_search_if} is not necessary, but it prevents us from wasting resources on computing the gradient when the energy is anyway changing drastically.
Although in our examples the additional cost of the gradient computation is small when amortized over all simulation steps, an application could use custom rules so that the approximate Armijo condition is evaluated more sparingly --- for example, only after a fixed number of Newton iterations or if the standard line search returns a small value of $\alpha$.

Revisiting the bottom plot in \figref{fig:robust_line_search_example}, we see that while the backtracking line search delivers a noisy approximation of $\Delta E$ that does not satisfy the Armijo condition, both $\Delta E_\text{approx}$ and our conservative estimate $\Delta E_\text{approx} + |\varepsilon_\text{est}|$ are smooth and satisfy sufficient descent.

We conclude by pointing out that we are unable to formally prove the robustness of our algorithm.
Doing so would likely require very particular assumptions on the form of the numerical errors in computing $E$, which may be manifold.
However, our proposed line search modification completely eliminated spurious line search failures in all our simulations.
We show examples of experiments that fail with the standard backtracking line search in \sectionref{sec:evaluation}.
\section{Convergence criteria}
\label{sec:convergence_criteria}
In this section, we first discuss existing choices of convergence criteria.
Our goal is to contribute some nuance that may help practitioners decide on an appropriate criterion for their application.
We finally propose a new acceleration-based criterion in  \sectionref{sec:acceleration_balance}, which rescales the residual vector into a more interpretable representation.

\subsection{Residual norm}
We are looking for solutions to the nonlinear equations \eqref{eq:backward_euler_residual}, and would like to terminate the optimization process once we are sufficiently close to a solution.
The classical example of a convergence criterion is to iterate until the residual norm satisfies $\| \nabla E\| = \| \vec r \| \leq \epsilon$ for some $\epsilon > 0$.

Finding an appropriate value for $\epsilon$ is however strongly problem-dependent.
For example, for finite element problems, each component in the residual corresponds to a node in the mesh, and the value is roughly proportional to the volume associated with the node.
To automatically determine an appropriate value, \citet{ZBK18} proposed a characteristic scaling factor of the residual norm for static equilibrium problems in geometry processing, which was later extended to incremental potentials by \citet{LGL+19}.
Their focus is on a visual error metric, which corresponds roughly to the displacement error $\| \vec u^* - \vec u\|$, where $\vec u^*$ is an exact solution.
For stiff materials, small displacements that are barely noticeable may still yield large residuals.
By design, their tolerance is therefore looser for increased stiffness, so that the displacement error may hopefully be held more or less constant.
However, the definition of the scaling factor assumes linear tetrahedral elements and a uniform material; a single material model and a single set of material parameters.
It is therefore unclear how to employ this convergence criterion for more general multibody simulations.

Since the residual is a measure of balance of forces, a simpler and more general alternative is to scale by a \emph{reference force} $\vec f_\text{ref}$~\cite{LLK+20}.
The convergence criterion then requires that $\| \vec r \| \leq \epsilon \| \vec f_\text{ref} \|$.
An example for the reference force can be the gravity force.
This choice does not incorporate stiffness in the scaling factor, so it will enforce smaller displacement errors for stiff materials.
This is, however, desirable for engineering, fabrication and some robotics applications where the approximation of the \emph{force} is of utmost importance.
To our knowledge, the effectiveness of this kind of scaling factor compared to the characteristic scaling factor of \citet{ZBK18} and \citet{LGL+19} has not been assessed experimentally.

Regardless of the scaling factor employed, a drawback of residual norm criteria is that they attempt to characterize the entire simulation by a single number.
Let us revisit the example of the two boxes from \figref{fig:box_example}.
If we consider the reference-based criterion, then $\vec f_\text{ref} = (\vec f^A_\text{ref}, \vec f^B_\text{ref})$.
If we assume that $\norm{\vec f^A} \ll \norm{\vec f^B}$ since the second body is much larger, we have that
\begin{align}
\|\vec f_\text{ref}\| = \sqrt{\|\vec f^A_\text{ref}\|^2 + \| \vec f^B_\text{ref}\|^2} \approx \|\vec f^B_\text{ref}\|.
\end{align}
Since $B$ is assumed to be in equilibrium, we have $\vec r^B = 0$, and the convergence criterion becomes
\begin{align}
\| \vec r \| =
\| \vec r^A \|
\leq \epsilon \| \vec f_\text{ref} \|
\approx \epsilon \| \vec f^B_\text{ref} \|.
\end{align}
This shows that the \emph{mere existence} of object $B$ has replaced the convergence criterion for object $A$ with a much looser tolerance, even though there is no interaction between them.
This non-local effect makes residual norm criteria difficult to reason about, especially in complex multibody multi-material simulations.

\subsection{Newton step length}
\label{sec:newton_step_length}
The residual norm essentially measures how well the forces are balanced.
For many computer graphics applications, only the visual outcome of the simulation matters, and so it makes sense to try to more directly approximate the displacement error.

\citet{LFS+20} suggest to use the max-norm of the Newton step scaled by the time step as a measure of sufficient convergence, i.e. $\| \Delta \vec u \|_\infty \leq \Delta t \, \epsilon_d$, in their Incremental Potential Contact (IPC) method, citing the natural scaling of the Hessian as a motivating factor.
This ensures that the maximum change to displacement of any individual node is below a certain threshold.

Although monitoring the step length is in general prone to mistaking stagnation for convergence, the exact Newton step nevertheless has a close connection to the \emph{actual} displacement error~\cite{Kel03}.
Under some regularity assumptions, we have that the exact Newton step $\Delta \vec u$ satisfies
\begin{align}
\Delta \vec u = \vec e + \mathcal{O} \big(\norm{\vec e}^2 \big),
\end{align}
which means that it is a first-order estimate of the error $\vec e = \vec u^* - \vec u$ (see \appendixref{sec:newton_first_order_estimate}).
For stiff and challenging problems, the constant terms hidden in the higher-order terms may nevertheless be significant, so that $\Delta \vec u$ may only be a reliable indicator of error when $\Delta \vec u$ is truly small.
The regularity assumptions may also not hold for many practical problems.

Care must also be taken if $\Delta \vec u$ is only an \emph{inexact} Newton step, e.g. stemming from an iterative solution procedure.
Furthermore, the Newton directions used in IPC stem from Projected Newton, i.e. $\vec H_\text{proj}$ is used in place of $\vec H$, in which case the direction is no longer a first-order estimate of the true error.
In our experience, the steps generated with $\vec H_\text{proj}$ are often substantially shorter than the full Newton step, which begs the question as to how reliable monitoring the Projected Newton step lengths is for true convergence.
At the same time, it avoids the problem of singular Hessians, since if $\vec H$ is nearly singular, the computed Newton step may be very small even if the error is large.

Monitoring the step length works particularly well for very stiff problems if only displacement error matters.
For example, the contact barriers used in IPC are very sensitive, in the sense that --- depending on the choice of the distance threshold $\hat{d}$ in the IPC model --- micrometer steps may result in comparatively large changes to force magnitudes.
Hence, a residual norm criterion might therefore insist on many additional iterations that do not change the visual outcome of the simulation.
Since we compare the max norm against a constant that only depends on the time step, every node must independently satisfy the criterion, so it also does not suffer from the non-local problems of the residual criterion.
However, it cannot be used with first-order methods where the operator $\vec H^{-1}$ is not available.

Finally, a significant practical drawback to monitoring the Newton step length is that it assumes that the Newton step is computed correctly in the first place.
As a result, implementation bugs for example in the assembly of $\vec H$ may more easily go undetected, which may otherwise manifest themselves as non-convergence under a residual criterion.

\subsection{Acceleration balance}
\label{sec:acceleration_balance}
We have seen in \sectionref{sec:newton_step_length} how rescaling the residual vector with $\vec H^{-1}$ allows us to use the max norm, which avoids the non-local problems of the residual norm criterion.
This rescaling may be suitable for exact Newton methods when displacement error is the only metric.
If \emph{actual} force balance is needed, one would have to carefully adjust $\epsilon_d$ based on the stiffness of the material.
Additionally, if $\epsilon_d$ is chosen too coarsely, it may still mistake intermittent shorter steps for convergence.

This motivates us to propose an alternative rescaling of the residual equations that is more locally interpretable than the residual equations, in the sense that it can be understood in a point-wise manner at any point in our domain, just like the Newton step length criterion.
This is not the case for the residual, where the residual force associated with each node depends on its volume.
Rescaling to force density would require problem-specific quantities not readily at hand.

The residual \eqref{eq:backward_euler_residual} is a direct numerical discretization of the equations of motion, which measures the balance of forces and inertia.
We denote the acceleration by $\vec a = \frac{\vec u - \tilde{\vec u}}{(\Delta t)^2}$ and represent the residual \eqref{eq:backward_euler_residual} as $\vec r = \vec M \vec a - \vec f$.
We can rewrite the balance of forces as \emph{balance of acceleration} by multiplying with the inverse mass, so that we obtain the \emph{acceleration residual} $\vec r_a$:
\begin{align}
\vec r_a := \vec a - \vec M^{-1} \vec f = \vec M^{-1} \vec r.
\end{align}
This is a measure of how well the acceleration induced by applied forces explain the currently observed acceleration.
Since acceleration is a field quantity, the residual acceleration is well-defined at any point in a continuum domain, and we can directly look for the maximum residual acceleration among all of our nodes
\footnote{
Because linear elements are interpolatory, this is exactly the maximum of all points in our discretization.
For higher-order elements, this is an approximation.
}, which motivates us to propose the convergence criterion
\begin{align*}
\| \vec r_a \|_\infty \leq \epsilon_a,
\end{align*}
where $\epsilon_a$ is given in units of acceleration.
For our experiments, we found that the values of $\epsilon_a = \SI{1.0}{\meter \per \second \squared}$ and $\epsilon_a = \SI{0.01}{\meter \per \second \squared}$ served well as examples of respectively \emph{coarse} and \emph{fine} tolerances across all of our examples.
The coarse tolerance was picked because setting it much higher would sometimes lead to unsatisfied boundary conditions with the penalty method.
The results of the coarse and fine tolerances mostly looked visually identical.
For brevity, we will omit units in the rest of the paper.

If $\vec M$ is not diagonal, computing $\vec r_a$ has non-negligible cost.
We have found that using $\operatorname{diag}(\vec M)^{-1}$ in place of $\vec M^{-1}$ introduces only a small error which can be compensated for by choosing a slightly smaller value of $\epsilon_a$.
The discrepancy is small for linear elements, but increases  slightly with element order.
For our results, we use a sparse Cholesky factorization for $\vec M^{-1}$, pre-computed at the beginning of each time step.

\subsection{Summary}
The right choice of convergence criterion depends on the problem.
We believe that our acceleration-based criterion may be a good alternative to appropriately scaled residual norm criteria since it does not suffer from the non-local effects of the residual norm.
The Newton step length criterion is particularly useful in situations where the residual is extraordinarily sensitive and displacement error is the primary metric.
We use both criteria in our experiments in \sectionref{sec:evaluation}.
\section{Implementation}
\label{sec:implementation}
We briefly outline some high-level implementation choices that impact all of our experiments.
\subsection{FEM}
We use the consistent mass matrix for all simulations, i.e. no mass lumping, and state-of-the-art analytic semidefinite projection~\cite{KE22} unless otherwise stated.

We consider two different methods for handling simple boundary constraints: \emph{direct imposition} and \emph{penalty}. We will see in \sectionref{sec:evaluation_elasticity} that the boundary handling approach may strongly impact convergence.
Details of the implementation may be found in \appendixref{sec:implementation_boundary_constraints_details}.

\subsection{IPC contacts}

We use the \texttt{IPC Toolkit}~\cite{ipc_toolkit} for integrating the IPC frictional contact model with our solvers.
We use its area-weighted barrier formulation for more consistent behavior across mesh resolutions.

We have found the results of the IPC simulations to be non-deterministic for challenging problems, particularly with friction.
The iteration counts can sometimes vary by a significant amount (typically up to around 10\%, but we have observed more than 50\% deviation in extreme cases not included in the paper).
The issue here is not actually the non-determinism, which stems from parallelism.
The fact that we can observe such large differences between runs of the same simulation means that the results are highly sensitive to perturbations.
Therefore, we advise readers to focus on trends across experiments rather than exact iteration numbers.

The original IPC algorithm uses an automatic selection and adaptive update strategy for the \emph{barrier stiffness} $\kappa$.
Since changing $\kappa$ changes the incremental potential itself, which would make our results harder to interpret and analyze, we elect to keep $\kappa$ fixed, and select an appropriate value for each experiment.
\section{Evaluation}
\label{sec:evaluation}

  \newcommand{\addlabel}[4]{
    \begin{minipage}[t][#3][c]{\linewidth}
	  \vspace*{\fill}
      #1
	  \vspace*{\fill}
      \caption{#2.}
      \label{#4}
    \end{minipage}
  }

\begin{figure}[tb!]
  \centering

  \begin{subfigure}{0.48\columnwidth}
    \addlabel{\includegraphics[width=\linewidth, trim = {0, 0.5cm, 0, 2cm}, clip]{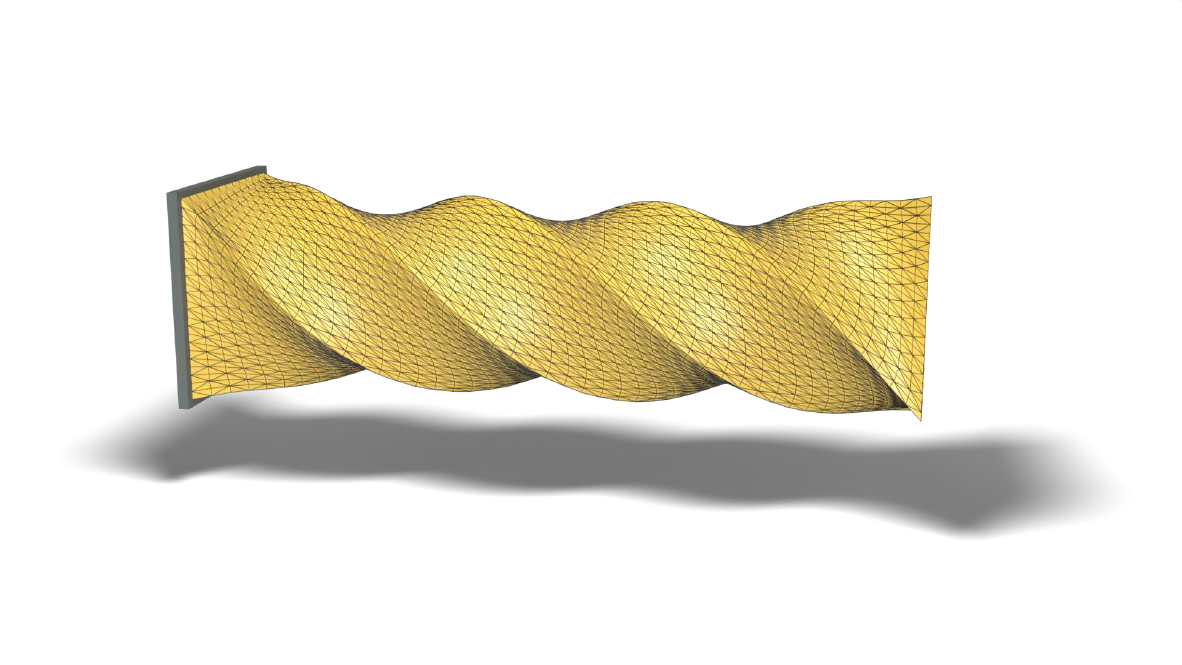}}{Twisting Beam}{3cm}{fig:twisting_beam_render}
  \end{subfigure}%
  \begin{subfigure}{0.48\columnwidth}
    \addlabel{
	\begin{tikzpicture}
      \node[anchor=south west, inner sep=0] (image2) at (0,0) {\includegraphics[width=\linewidth, trim ={0, 0.5cm, 0, 0cm}, clip]{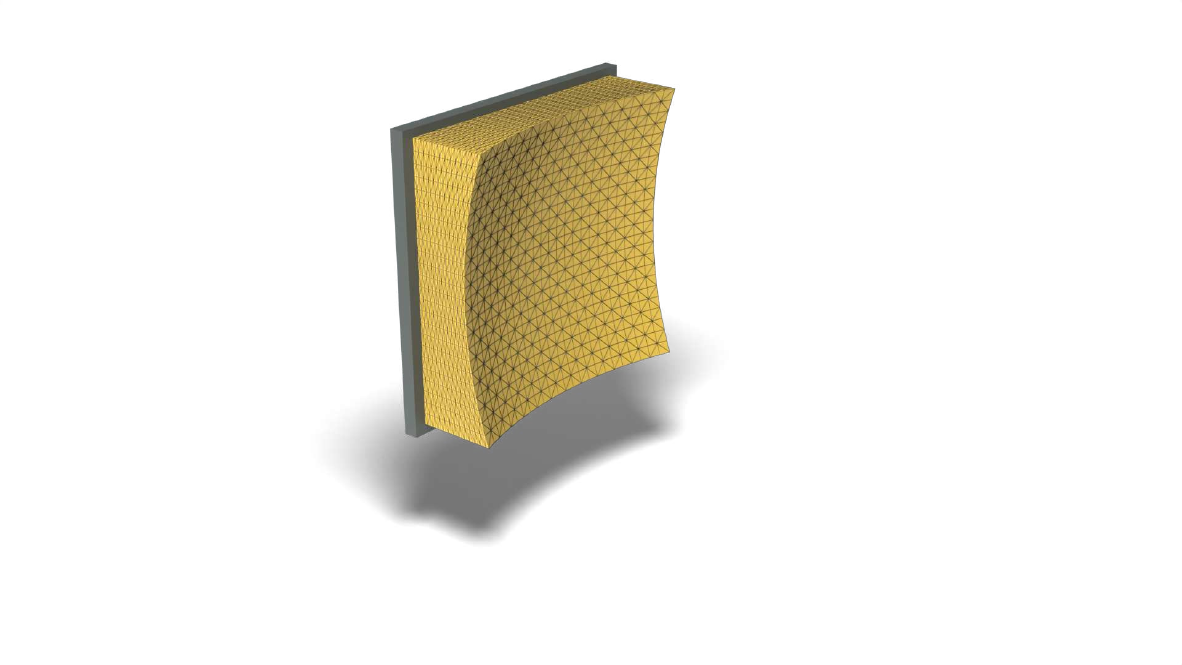}};
      \node[anchor=south west, inner sep=0, opacity=0.15] (image1) at (0,0) {\includegraphics[width=\linewidth, trim={0, 0.5cm, 0, 0cm}, clip]{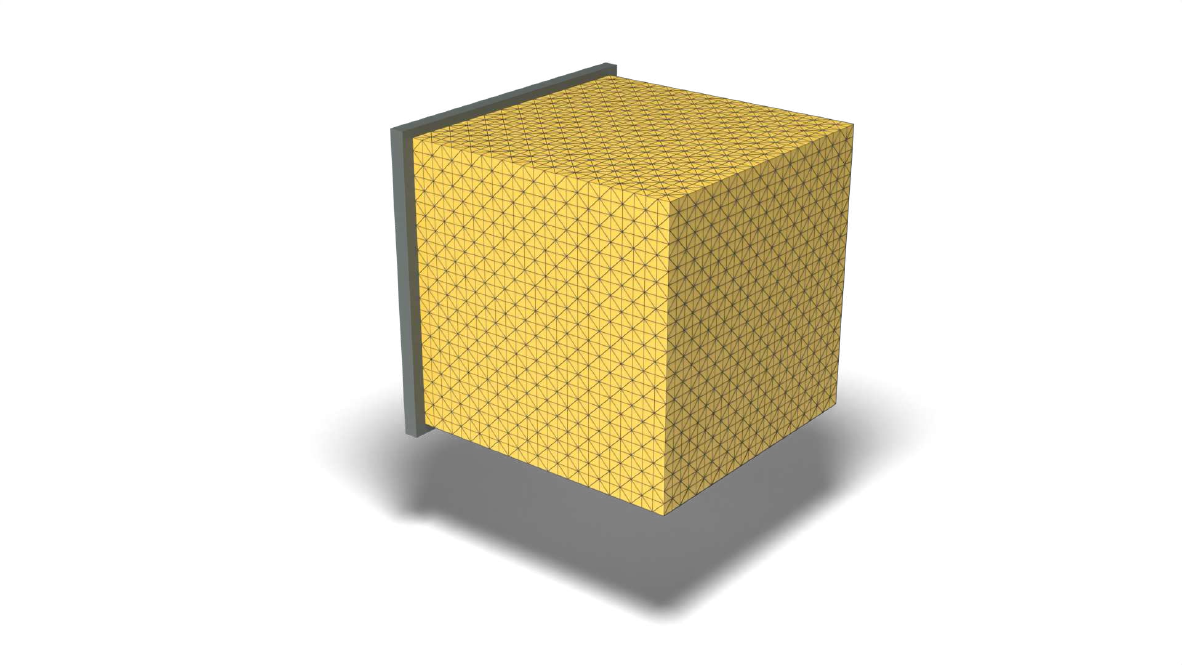}};
    \end{tikzpicture}
    }{Compressing Box}{3cm}{fig:compressing_box_render}
  \end{subfigure}

  \begin{subfigure}{0.48\columnwidth}
    \addlabel{\includegraphics[width=\linewidth, trim = {1cm, 1cm, 2cm, 4cm}, clip]{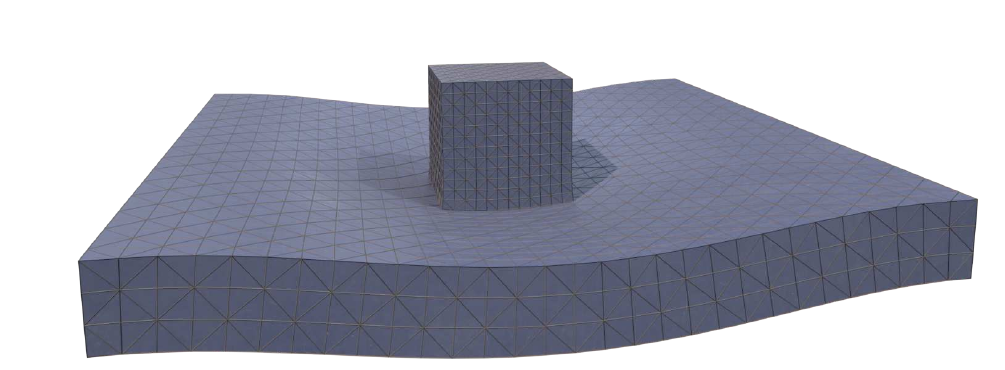}}{Basic Impact (Soft)}{2.5cm}{fig:basic_impact_soft_render}
  \end{subfigure}%
  \begin{subfigure}{0.48\columnwidth}
    \addlabel{\includegraphics[width=\linewidth, trim = {1cm, 1cm, 2cm, 4cm}, clip]{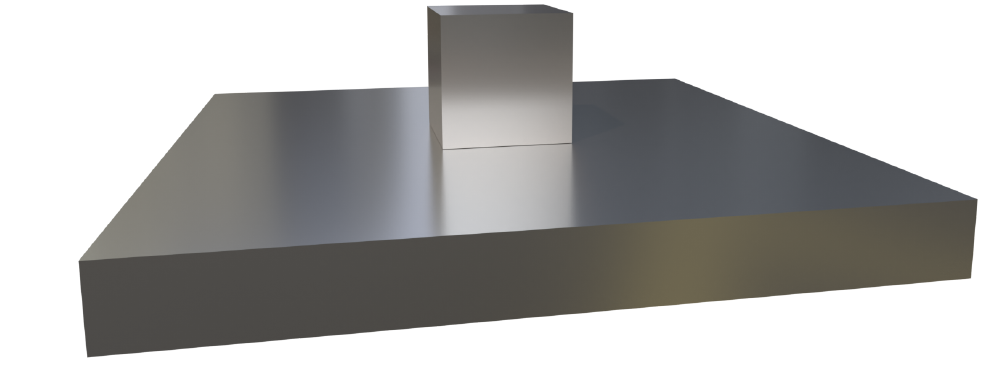}}{Basic Impact (Stiff)}{2.5cm}{fig:basic_impact_stiff_render}
  \end{subfigure}

  \caption{Snapshots of some of the synthetic experiments used for the evaluation.
  The transparent box for the Compressing Box indicates the initial state.
  }
\end{figure}

\begin{figure*}[tb!]
  \begin{subfigure}{0.39 \textwidth}
    \addlabel{\includegraphics[width=\linewidth, trim = {1.5cm, 2.3cm, 1.5cm, 3cm}, clip]{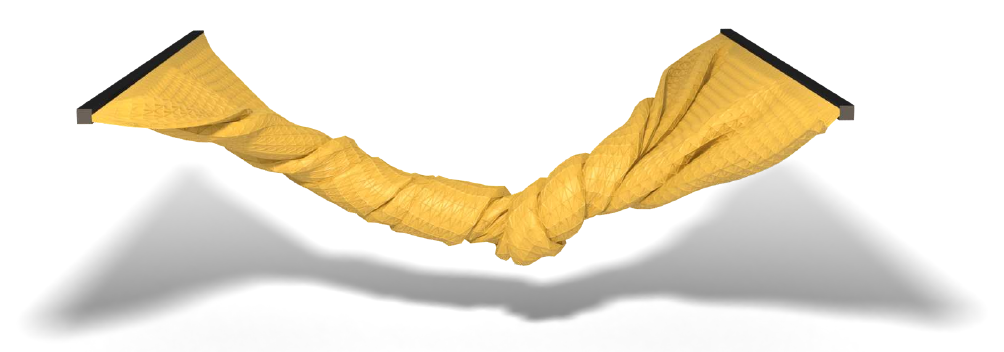}}{Twisting Mat}{5.5cm}{fig:twisting_mat_render}
  \end{subfigure}
  \begin{subfigure}{0.3 \textwidth}
    \addlabel{\includegraphics[width=\linewidth, trim = {0, 0.5cm, 6.5cm, 0.2cm}, clip]{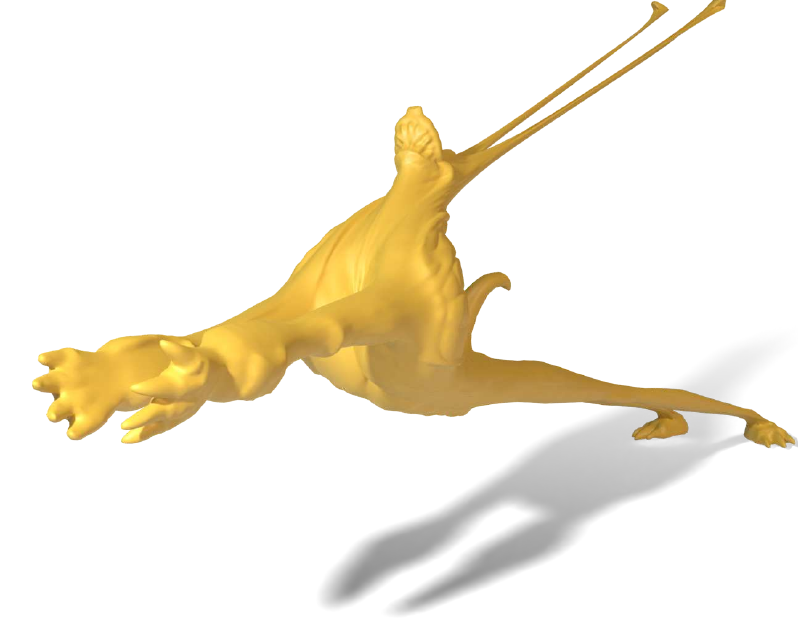}}{Armadillo Slingshot}{5.5cm}{fig:armadillo_slingshot_render}
  \end{subfigure}%
  \begin{subfigure}{0.3 \textwidth}
    \addlabel{\includegraphics[width=\linewidth, trim = {4cm, 1.1cm, 4cm, 0.7cm}, clip]{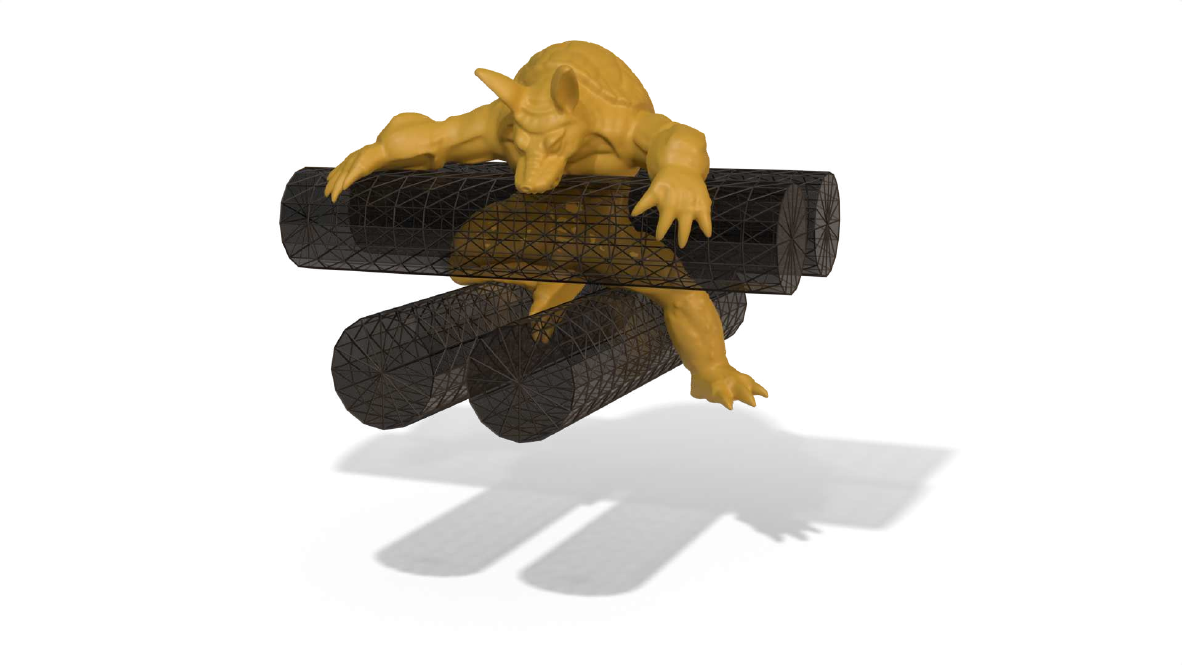}}{Armadillo Rollers}{5.5cm}{fig:armadillo_rollers_render}
  \end{subfigure}%

  \caption{Snapshots of some of the experiments used for the evaluation
  }
  \label{fig:grid}
\end{figure*}

We evaluate the four solvers Newton (N), Projected Newton (PN), POD-Newton (PDN) and Kinetic Newton (KN) in a series of experiments.
The experiments are designed to stress the solvers in different ways.
The results then give us insights into the strengths and weaknesses of each solver.
We focus primarily on \emph{iteration count}, as this is the best measure we have of step quality.
We include representative examples of the experiments in the supplemental video.
These examples are intended to give the reader a clear idea of the setup and the dynamics of each experiment, but not to be exhaustive; the sheer number of simulations required for the results makes presenting all simulations impractical.

\begin{table*}[htb]
\caption{Parameters used for experiments: Duration $T$, time step $\Delta t$, Young's modulus $E$, Poisson's ratio $\nu$, density $\rho$, and IPC parameters $\mu$ (coefficient of friction), $\hat{d}$ (distance threshold) and $\epsilon_v$ (static friction accuracy), and finally penalty parameter $\sigma$ where applicable.
Experiments without contact forces do not have values for contact-related parameters.
}
\label{tab:parameters}

\rowcolors{2}{table_alternate}{white}
\begin{tabular}{lrrrrrrrrr}
\toprule
Experiment
	& $T$ (\si{\second})
	& $\Delta t$ (\si{\milli\second})
	& $E$ (\si{\pascal})
	& $\nu$
	& $\rho$ \big(\si[per-mode=fraction]{\kilo\gram\per\cubic\meter}\big)
	& $\mu$
	& $\hat{d}$ (\si{\meter})
	& $\epsilon_v$ (\si[per-mode=symbol]{\meter\per\second})
	& $\sigma$ (\si{\second^{-2}})
	\\
\midrule
Swinging Beam
	& $6.0$
	& $16.7$
	& $4 \cdot 10^5$
	& $0.40$
	& $1000$
	& ---
	& ---
	& ---
	& ---
	\\
Twisting Beam (moderate $\Delta t$)
	& $3.0$
	& $33.3$
	& $10^7$
	& $0.49$
	& $1000$
	& ---
	& ---
	& ---
	& $10^8$
	\\
Twisting Beam (large $\Delta t$)
	& $3.0$
	& $333$
	& $10^7$
	& $0.49$
	& $1000$
	& ---
	& ---
	& ---
	& $10^8$
	\\
Compressing Box
	& $6.0$
	& $10.0$
	& $10^5$
	& $0.40$
	& $1000$
	& ---
	& ---
	& ---
	& $10^{10}$
	\\
Armadillo Slingshot
	& $8.0$
	& $16.7$
	& $8 \cdot 10^4$
	& $0.40$
	& $1000$
	& ---
	& ---
	& ---
	& $10^8$
	\\
Basic Impact (soft)
	& $1.5$
	& $16.7$
	& $10^6$
	& $0.40$
	& $1000$
	& $0.1$
	& $10^{-3}$
	& $10^{-3}$
	& ---
	\\
Basic Impact (stiff)
	& $0.17$
	& $16.7$
	& $220 \cdot 10^{9}$
	& $0.30$
	& $8000$
	& $0.1$
	& $10^{-3}$
	& $10^{-3}$
	& ---
	\\
Twisting Mat
	& $15.0$
	& $16.7$
	& $2.0 \cdot 10^4$
	& $0.40$
	& $1000$
	& $0.0$
	& $1.4 \cdot 10^{-3}$
    & ---
    & $10^8$
	\\
Armadillo Rollers
	& $8.0$
	& $16.7$
	& $5.0 \cdot 10^4$
	& $0.20$
	& $1000$
	& $0.5$
	& $10^{-3}$
	& $10^{-3}$
	& $10^8$
	\\
Friction Twist (rubber structure)
	& $10.0$
	& $16.7$
	& $10^7$
	& $0.49$
	& $1300$
	& $1.0$
	& $10^{-3}$
	& $10^{-3}$
	& $10^9$
	\\
Friction Twist (steel plate)
	&
	&
	& $220 \cdot 10^9$
	& $0.30$
	& $7850$
	&
	&
	&
	&
	\\
\bottomrule
\end{tabular}
\end{table*}

\subsection{Setup}

All simulations were run on a 32-core AMD Ryzen Threadripper PRO 5975WX CPU with 256 \si{\giga \byte} RAM.
Our implementation is written in Rust.
We use \texttt{nalgebra}~\cite{nalgebra} and a variety of custom routines for basic dense and sparse linear algebra, but use  \texttt{Intel oneMKL 2023}~\cite{onemkl} for all sparse direct solvers, which we use for solving the linear system associated with each solver step.

Unless otherwise noted, we use direct imposition for static boundary conditions and penalties to impose moving boundary conditions.
We use linear tetrahedral elements for all simulations except where otherwise noted.
Experiment parameters such as time step, material and IPC model parameters are listed in \tableref{tab:parameters}.

Our robust line search succeeded across all experiments, but the standard backtracking line search failed for some, which we explicitly discuss. We declare a failure if $\alpha < 10^{-7}$ during line search.

\subsection{Elasticity}
\label{sec:evaluation_elasticity}

\subsubsection{Swinging beam}
Rerunning the Swinging Beam experiments with POD-Newton and Kinetic Newton, we verify that POD-Newton and Kinetic Newton take the same number of iterations as Newton.
This demonstrates that both methods reduce to Newton's method for sufficiently ``simple" problems, which was a key motivation behind their designs.

\subsubsection{Twisting Beam}
\label{sec:twisting_beam}
We stretch and twist a beam for $3$ seconds (\figref{fig:twisting_beam_render}). We consider two different regular spatial discretizations with $17.7 \si{k}$ (\emph{coarse}) and $58.2 \si{k}$ (\emph{fine}) nodes, and two time steps $\Delta t = 1/30 \si{\second}$ (\emph{moderate}) and $\Delta t = 1/3 \si{\second}$ (
\emph{large}).
\begin{table}[tb!]
\caption{Average number of iterations per time step for different configurations of the \emph{Twisting Beam} experiment.
Bold and underlined numbers indicate the smallest number of iterations (or within 5\%). \texttt{\#V} indicates the number of vertices in the mesh.}
\label{tab:twisting_beam}
\rowcolors{2}{table_alternate}{white}
\begin{tabular}{ccccrrrr}
\toprule
\cellcolor{white}  \multirow{2}{*}{$\Delta t$}\cellcolor{white}  & \multirow{2}{*}{\#V}\cellcolor{white}  & \multirow{2}{*}{$\epsilon_a$}\cellcolor{white}  & \multicolumn{4}{c}{Iterations per step} \\
\cmidrule{4-7}
\cellcolor{white} & \cellcolor{white} & \cellcolor{white} & \cellcolor{white}N & \cellcolor{white}PN & \cellcolor{white}PDN & \cellcolor{white}KN \\
\midrule
33.3 \si{\milli \second} & 17.7k & 0.01 & $\underline{\mathbf{4.4}}$ & $21.9$ & $\underline{\mathbf{4.4}}$ & $\underline{\mathbf{4.4}}$\\
 &  & 1.00 & $\underline{\mathbf{3.1}}$ & $13.8$ & $\underline{\mathbf{3.1}}$ & $\underline{\mathbf{3.1}}$\\
 & 58.2k & 0.01 & $\underline{\mathbf{4.1}}$ & $19.5$ & $\underline{\mathbf{4.1}}$ & $\underline{\mathbf{4.1}}$\\
 &  & 1.00 & $\underline{\mathbf{3.4}}$ & $12.5$ & $\underline{\mathbf{3.4}}$ & $\underline{\mathbf{3.4}}$\\
333 \si{\milli \second} & 17.7k & 0.01 & $141.8$ & $52.0$ & $\underline{\mathbf{10.9}}$ & $15.2$\\
 &  & 1.00 & $134.8$ & $34.9$ & $\underline{\mathbf{10.5}}$ & $14.4$\\
 & 58.2k & 0.01 & $210.9$ & $46.0$ & $\underline{\mathbf{13.2}}$ & $16.7$\\
 &  & 1.00 & $209.1$ & $31.3$ & $\underline{\mathbf{11.7}}$ & $16.0$\\
\bottomrule
\end{tabular}
\end{table}
~ 
\begin{figure}[htb]
    \begin{center}
        \includegraphics[width=\columnwidth]{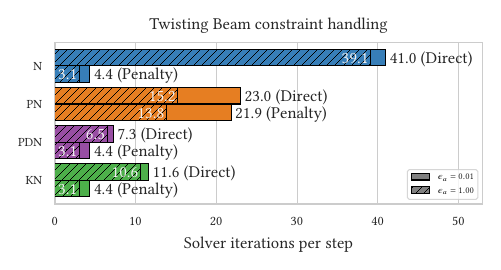}
    \end{center}
    \caption{Effect on total solver iteration count due to choice of constraint enforcement method for the twisting beam experiment with \emph{coarse} spatial resolution and \emph{moderate} time step.}
    \label{fig:twisting_beam_constraint_enforcement_bar_plot}
\end{figure}

\paragraph{Impact of boundary handling method}
We first study the effect the choice of boundary handling method has on the solver performance.
We run simulations with both direct imposition (\emph{Direct}) and the penalty method (\emph{Penalty}) for boundary conditions and compare the results, which are summarized in \figref{fig:twisting_beam_constraint_enforcement_bar_plot}.
The Direct simulations failed for the fine mesh or the large time step, since the direct imposition resulted in an invalid (inverted) state with infinite energy.
Therefore we only consider results for the coarse mesh and moderate time step for this comparison.

Here we first observe that POD-Newton, Newton and Kinetic Newton all behave identically for Penalty, and require only a small additional number of iterations to reach the stricter convergence criterion.
Projected Newton, in comparison, requires far more iterations, and the gap between the coarse and fine convergence criteria is once again substantially larger, indicating a reduced local convergence rate for Projected Newton.

The difference in performance between Direct and Penalty for Newton's method is stark and surprisingly severe. Newton requires more than 9x iterations for Direct compared to Penalty.
In contrast, POD-Newton suffers far less from direct imposition, requiring about 1.7x-2x iterations compared to Penalty.
Here, the semidefinite projection is clearly responsible for getting out of the difficult state that the direct imposition has put the simulation into, which eventually enables full Newton steps to be effective.
Kinetic Newton clearly outperforms Newton in this context, but it takes 1.6x iterations compared to POD-Newton.
This suggests that semidefinite projection is more effective than our regularization scheme at getting out this type of difficult intermediate states.
Finally, the number of iterations for Projected Newton does not change substantially with boundary handling method, likely because its iteration counts are already dominated by comparatively slow local convergence.

\begin{figure*}[htb]
	\centering
	\begin{minipage}{\columnwidth}
		\centering
		\includegraphics[width=\columnwidth]{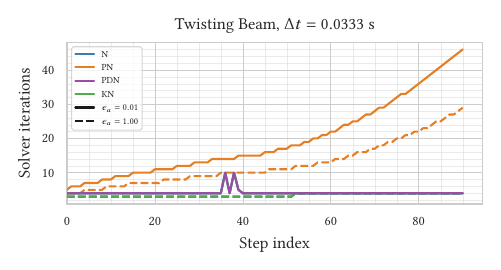}
	\end{minipage}
	\hfill
	\begin{minipage}{\columnwidth}
		\centering
		\includegraphics[width=\columnwidth]{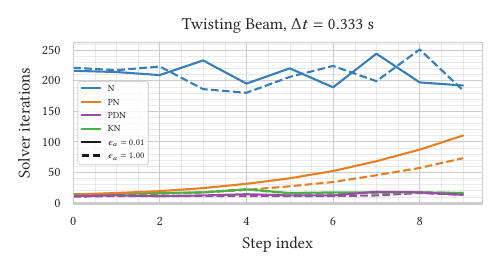}
	\end{minipage}
	\caption{Iteration counts per time step for the ``fine" Twisting Beam experiment for both the moderate and large time step.}
	\label{fig:twisting_beam_per_step_counts}
\end{figure*}

\paragraph{Per-step iteration counts}
We show the iteration count per step for a simulation with the fine mesh, for both moderate and large time steps, in \figref{fig:twisting_beam_per_step_counts}.
Iteration counts are additionally given in \tableref{tab:twisting_beam}.
As before, the iteration count for PN increases in tune with the severity of the deformation.
For the moderate time step, N, KN and PDN all behave identically, and enjoy an almost constant iteration count per step.
The situation is different for the large time step, where we see that Newton requires an order of magnitude or more iterations than PDN and KN.
PDN once again comes out on top, with KN close behind with 1.3x-1.4x iterations compared to PDN.
~
\subsubsection{Compressing box}
\label{sec:compressing_box}
~
\begin{figure}[htb]
    \begin{center}
        \includegraphics[width=\columnwidth]{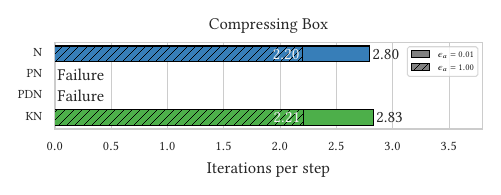}
    \end{center}
    \caption{Average solver iterations per time step for the \emph{compressing box} experiment (\sectionref{sec:compressing_box}).}
    \label{fig:compressing_box_bar_plot}
\end{figure}
~
We constrain \emph{all} sides of a box with moving boundary conditions so that only the interior can deform (\figref{fig:compressing_box_render}).
One side remains static for the duration of the simulation, whereas the other five sides deform.
We proceed to gradually squash the box towards a nearly flat configuration (see supplemental video).
The displacement of the side opposite to the static side is non-uniform in order to make the resulting internal deformation non-trivial.
The mesh has 29.4k vertices and 165.8k elements.
\figref{fig:compressing_box_bar_plot} demonstrates the average number of solver iterations per time step for each solver.
Notably, Projected Newton and POD-Newton failed because they both exceeded our upper limit of 1000 solver iterations in a single time step, whereas Newton and Kinetic Newton behave very similarly.
Neither seems to have particular trouble converging for this numerically challenging example.

This shows that, at least in rare or contrived cases, the semidefinite projection mechanism can fail to make meaningful progress, leading to convergence failure.

We additionally ran all simulations for this experiment with the standard backtracking line search.
The standard line search failed in all cases, thereby halting the simulation.
In contrast, our robust line search enabled the optimization process to continue.

\subsubsection{Armadillo Slingshot}
\begin{figure}[htb]
    \begin{center}
        \includegraphics[width=\columnwidth]{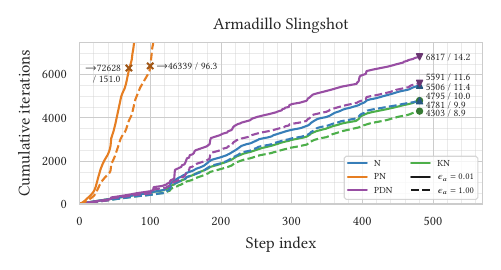}
    \end{center}
    \caption{Cumulative solver iterations per time step for the Armadillo Slingshot experiment.
	The total number of solver iterations and average number of iterations per time step are annotated next to the curve for each solver.
    }
    \label{fig:armadillo_slingshot_cumulative}
\end{figure}
~
Inspired by the Armadillo stress test devised by \citet{LGL+19}, we stretch a soft Stanford Armadillo, and release it after building up significant strain (\figref{fig:armadillo_slingshot_render}).
The mesh has 23.7k vertices and 97.1k elements.
\figref{fig:armadillo_slingshot_cumulative} depicts the \emph{cumulative} number of iterations per time step, since a plot of the iteration count per step would be very noisy.
The slope of this curve indicates how many iterations are needed per time step.
The curve for PN is very steep, indicating that it needs far more iterations per step, and finishes with an iteration count roughly an order of magnitude or greater compared to the other methods.
For this experiment, Newton performs very well, only beaten by KN.
POD-Newton's slightly worse performance could indicate that projection is less effective here, but it could also be random: the supplemental video shows that each solver takes a somewhat different trajectory, and some trajectories may be harder to solve than others.

\begin{table*}[htb!]
\centering
\caption{Timing breakdown for the Armadillo Slingshot experiment, for the coarse (top) and fine (bottom) convergence tolerances, in addition to the average time per iteration.
Percentages are relative to total time.
The best (within 5\%) numbers for each convergence tolerance is underlined and bold.}
\label{tab:armadillo_slingshot_timing}
\rowcolors{2}{table_alternate}{white}
\begin{tabular}{lrrrrrrrr}
\toprule
Solver & $\epsilon_a$ & \# Iter & \# Iter / step & Line search (s) & Assembly (s) & Linear solve (s) & Total time (s) & Avg. iter (s)\\
\midrule
Kinetic Newton & 1.00 & \underline{\textbf{4303}} & \underline{\textbf{8.9}} & \underline{\textbf{ 52}} (3.4\%) & \underline{\textbf{ 99}} (6.4\%) & 1322 (84.7\%) & 1561 & 0.363\\
Newton & 1.00 & 4781 & 9.9 &  85 (5.8\%) & 121 (8.3\%) & \underline{\textbf{1190}} (81.5\%) & \underline{\textbf{1459}} & \underline{\textbf{0.305}}\\
POD-Newton & 1.00 & 5591 & 11.6 &  66 (3.6\%) & 170 (9.2\%) & 1498 (81.6\%) & 1836 & 0.328\\
Projected Newton & 1.00 & 46339 & 96.3 & 506 (3.6\%) & 1323 (9.4\%) & 11653 (82.9\%) & 14055 & \underline{\textbf{0.303}}\\
\midrule
Kinetic Newton & 0.01 & \underline{\textbf{4795}} & \underline{\textbf{10.0}} & \underline{\textbf{ 64}} (3.7\%) & \underline{\textbf{119}} (6.9\%) & 1452 (83.9\%) & \underline{\textbf{1732}} & 0.361\\
Newton & 0.01 & 5506 & 11.4 & 102 (6.1\%) & 139 (8.3\%) & \underline{\textbf{1368}} (81.4\%) & \underline{\textbf{1681}} & \underline{\textbf{0.305}}\\
POD-Newton & 0.01 & 6817 & 14.2 &  85 (3.8\%) & 204 (9.1\%) & 1833 (81.5\%) & 2248 & 0.330\\
Projected Newton & 0.01 & 72628 & 151.0 & 960 (4.4\%) & 2065 (9.5\%) & 17864 (81.8\%) & 21842 & \underline{\textbf{0.301}}\\
\toprule
\end{tabular}

\end{table*}

We also provide a basic timing breakdown in \tableref{tab:armadillo_slingshot_timing},
where the time to check the convergence criterion has been subtracted from the total time.
While KN takes fewer iterations, Newton is slightly faster measured in time.
KN has the highest average iteration time, because it requires the most factorization attempts, and the simulation is dominated by the cost of the direct solver.
PDN also has a higher average iteration time than Newton and PN, presumably because like KN it must occasionally also suffer a failed factorization and an extra call to the assembly routine.

\subsection{IPC contacts}
\label{sec:evaluation_contact}
We evaluate the different solvers on examples with contact.
We use the state-of-the-art IPC model~\cite{LFS+20}, which represents frictional contact (with lagged friction) as a monolithic incremental potential.

The frictional IPC contact forces are very sensitive --- a tiny perturbation in displacement leads to a significant change in the applied contact force.
This means that very small displacements are needed to reach force or acceleration balance, sometimes on the order considerably smaller than nanometers, at which point numerical issues may cause problems for convergence.
For this reason, we use the step length criterion (\sectionref{sec:newton_step_length}) for all IPC experiments, with a \emph{coarse} tolerance $\epsilon_d = 10^{-2} \si{\meter\per\second}$ and a \emph{fine} tolerance $\epsilon_d = 10^{-4} \si{\meter \per \second}$.

\subsubsection{Basic impact}
\label{sec:basic_impact}
\begin{figure}[htb]
    \begin{center}
        \includegraphics[width=\columnwidth]{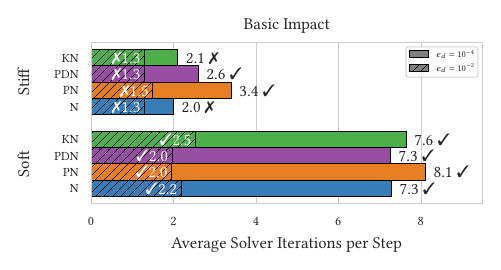}
    \end{center}
    \caption{Average solver iterations per time step for the \emph{basic impact} experiment (\sectionref{sec:basic_impact}), for both the \emph{stiff} and \emph{soft} material. Standard line search failure is indicated by a cross, success is indicated by a checkmark.}
    \label{fig:basic_impact_bar_plot}
\end{figure}
~
We let a box fall with high initial velocity ($\SI{10} {\meter\per\second}$) onto another cuboid whose sides are constrained to have zero displacement, leading to a high velocity impact.
The two meshes together contain 10.5k vertices and 53.1k elements.
We perform the experiment with both a soft (\figref{fig:basic_impact_soft_render}) and a stiff material (\figref{fig:basic_impact_stiff_render}).
We only record iterations up to the point where the simulation has visually settled, which is after 90 steps (\SI{1.5}{\second}) for the soft material, and 10 steps (\SI{167}{\milli\second}) for the stiff material.

The average number of solver iterations per step is presented in \figref{fig:basic_impact_bar_plot}.
Curiously, all four solvers seem to perform somewhat similarly here, with Projected Newton being almost on par with the other solvers.
We believe that the explanation is not that Projected Newton is particularly \emph{good}, but rather that all solvers perform rather poorly.
This is suggested by the significant difference between the performance for the \emph{coarse} and \emph{fine} convergence criteria.
This is in strong contrast to what we observed for the elasticity experiments, where the faster solvers would only require a small number of additional iterations to satisfy the \emph{fine} criterion compared to the \emph{coarse}.

\figref{fig:basic_impact_bar_plot} also shows that, while our robust line search succeeded in all cases, the standard backtracking line search failed for 6/8 simulation configurations for the \emph{stiff} material, although it succeeded for the \emph{soft} material.

\subsubsection{Twisting mat}
\begin{figure}[htb]
    \begin{center}
        \includegraphics[width=\columnwidth]{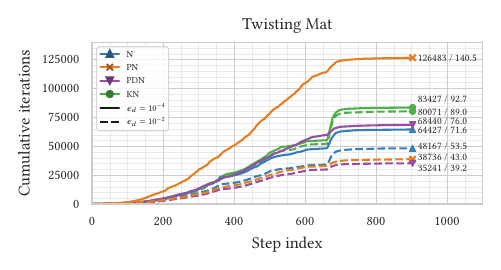}
    \end{center}
    \caption{Cumulative solver iterations per time step for the Twisting Mat experiment.
	The total number of solver iterations and average number of iterations per time step are annotated next to the curve for each solver.
    }
    \label{fig:twisting_mat_cumulative}
\end{figure}
~
Similar to the \emph{twisting mat} stress test used by \citet{LFS+20}, we twist a soft, thin mat for $11$ seconds ~(\figref{fig:twisting_mat_render}), but then release it at time step index 660. The mesh consists of 19.5k vertices and 76.8k elements.
The \emph{cumulative} number of iterations at each step is depicted in \figref{fig:twisting_mat_cumulative}.
Here we see that for the \emph{coarse} convergence tolerance, Newton, PN and PDN behave similarly up to the point where the mat is released, while KN is lagging behind.
The iteration count for both N and KN increases rapidly right after the mat is released, though considerably more so for KN.
As a result, PDN and PN arrive at a similar total iteration count, with Newton following behind at 1.4x the PDN iteration count.
For the \emph{fine} tolerance, we see, perhaps surprisingly, that Newton outperforms all other solvers just ahead of PDN, while PN requires significantly more iterations, and KN performs roughly like it did for the coarse tolerance.

\subsubsection{Armadillo rollers}
\begin{figure}[tb!]
    \begin{center}
        \includegraphics{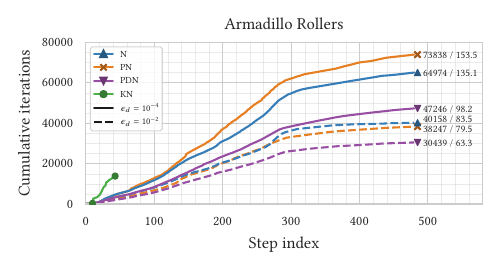}
    \end{center}
    \caption{Cumulative solver iterations per time step for the Armadillo Rollers experiment.
	The total number of solver iterations and average number of iterations per time step are annotated next to the curve for each solver.
	KN exceeded maximum iteration count in a single step.
    }
    \label{fig:armadillo_rollers_cumulative}
\end{figure}
~
We evaluate the solvers on a re-creation of the \emph{armadillo rollers} stress test demonstrated by \citet{LFS+20}, where a soft Stanford Armadillo falls through several hard cylindrical obstacles turning at a constant rate (\figref{fig:armadillo_rollers_render}).
The mesh has 23.7k vertices and 97.1k elements.
This experiment demonstrates strong frictional forces that drive complex deformation.
The cumulative solver iterations are depicted in \figref{fig:armadillo_rollers_cumulative}.
For the coarse tolerance, PDN requires the smallest number of iterations, though PN is not trailing far behind, followed by Newton.
For the fine tolerance, PDN is clearly superior while Newton and PN lag behind and behave somewhat similarly.
KN starts out poorly and is soon aborted due to reaching maximum iteration count for a single time step (6000).

\subsubsection{Friction Twist}
\begin{figure}[tb!]
    \begin{center}
        \includegraphics{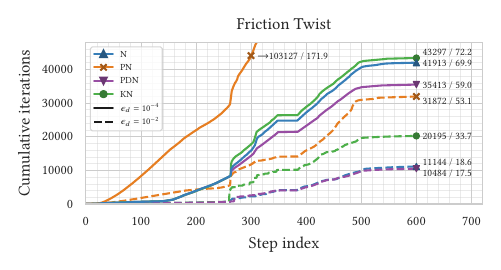}
    \end{center}
    \caption{Cumulative solver iterations per time step for the Friction Twist experiment.
	The total number of solver iterations and average number of iterations per time step are annotated next to the curve for each solver.
    }
    \label{fig:friction_twist_cumulative}
\end{figure}
~
We finally simulate a rubber-like, blue structure with cylindrical supports, standing on top of a static floor.
A steel plate with rotating boundary conditions for its top face gradually pushes down upon the structure.
The strong friction forces generated by the rotating steel plate cause the structure to deform --- as seen in \figref{fig:teaser} (Left) --- and eventually buckle.
Finally, the steel plate is released from its boundary constraints, causing it to slide off, which releases the elastic energy stored in the blue structure.

By looking at the cumulative iteration counts in \figref{fig:friction_twist_cumulative}, we see that PDN comes out on top, while PN once again lags behind Newton and PDN.
KN keeps up fairly well for the fine tolerance, and for the coarse tolerance it outperforms PN but takes noticeably more iterations compared to N and PDN.
We also observe that the \emph{slope} of the curves for KN is overall similar to that of N and PDN, but with a significant spike near steps 260-265, which is the moment when the structure buckles.

\subsection{Discussion}
\label{sec:evaluation_discussion}
The trend that emerges from all of our experiments is that Projected Newton often exhibits a severely reduced convergence rate compared to Newton's method when sufficiently close to the solution.
Even for our coarse stopping criteria, Projected Newton underperforms in almost all experiments.
Our experiments indicate that, among the Newton-type solvers we have considered, POD-Newton performs well most reliably.
These results together suggest that semidefinite projection should not be applied unconditionally.
It is an effective method for making rapid progress in early iterations when the incremental potential $E$ is strongly non-convex, but will otherwise slow convergence.

Kinetic Newton is largely competitive for problems involving only hyperelasticity, but struggles with challenging IPC contacts.
We believe that the primary difference between these two settings is a combination of smoothness and sensitivity.
A typical pattern that we see with Kinetic Newton as it encounters indefiniteness in problems involving only hyperelasticity is that it first requires a very small $\beta$ parameter, indicating strong regularization.
Subsequent steps then let $\beta$ smoothly increase, often all the way to $\beta = 1$, at which point we recover regular Newton iterations in a region of positive definiteness.
Sometimes we still run into an indefinite region and the process repeats.
For contact, we have observed that $\beta$ will be decreased and increased in a more abrupt, unpredictable pattern.
The IPC model only attains $C^1$ continuity, so its Hessian contribution is discontinuous.
Consequently, the \emph{definiteness} of the Hessian is also discontinuous.
Combined with the considerable sensitivity of IPC contact energies, we suspect that this leads to suboptimal convergence for all methods, but Kinetic Newton is hit particularly hard due to its reliance on a notion of smooth definiteness.
It is not clear to us why the Armadillo Rollers scene in particular is so problematic for Kinetic Newton.
To determine this, we might have to dive deep into the details of the \texttt{IPC Toolkit}, which is out of scope for our investigation.

It is also interesting to see how well Newton's method performs for most experiments, even for the challenging IPC contact scenarios.
While we are unable to give a satisfying explanation for why Newton is so surprisingly competitive for the contact experiments, we have observed that for these problems, a significant proportion of intermediate states $\vec u^k$ appear to have indefinite Hessians.
While positive definiteness is a sufficient condition for descent, it is not a necessary one.
It is therefore possible that Newton is able to find reasonable descent steps despite the Hessian being indefinite.
A second possible explanation is that \emph{all} solvers are doing rather poorly in the first place due to the strong sensitivity and lack of smoothness in the incremental potential, which perhaps acts as an equalizer.

Despite being aware of the deficiencies of the \emph{direct imposition} method of handling moving boundary conditions beforehand, we were still surprised by the severe effect on convergence in our experiments.
Our results suggest a strong preference for penalty-like handling of moving boundaries.
We have deliberately used a simple penalty approach in the interest of making comparisons easier to interpret.
We believe that the more sophisticated Augmented Lagrangian approach used by IPC~\cite[Technical Supplement]{LFS+20} should inherit the favorable effects that the penalty method appears to have on convergence.

Finally, we have observed how backtracking line search can fail particularly in scenarios with very stiff materials (such as steel) or in highly numerically challenging scenarios  where the system is very sensititive (like the compressing box).
Our proposed robust line search appears to reliably prevent breakdown of the line search when this occurs, eliminating the need for ad-hoc fallback solutions.
\section{Conclusion}
Our experimental study focuses on the Backward Euler incremental potential for deformable solids.
Because our observations are largely explained by innate mathematical properties of the Projected Newton method, we suspect that most of our findings will extend to other integrators and physical systems, but possibly with a varying degree of significance.
This would have to be experimentally validated.

To discern inherent properties of the Newton-type methods under consideration without having to consider the myriad of possible decisions involved in the design of an inexact Newton solver, we restricted the evaluation to use direct linear solvers.
This is not representative of many real-world problems, and even some of our own experiments would be more efficiently solved with an iterative solver.
While POD-Newton and Kinetic Newton both rely on Cholesky factorization as an indication of positive definiteness, they could for example be straightforwardly adapted to use the existence of negative curvature directions in a Newton-CG setting instead.
We encourage researchers to try this out for their own problems and report their findings.

Our IPC experiments and experience indicate that these problems rarely attain the fast Newton-style convergence that we associate with smooth problems.
It is not clear to us if this is inherent to the sensitivity of the contact problems themselves, or something that could be improved in the IPC model.
We hope our observations here can help motivate further research in this direction.

We know from the experiences of our colleagues that many researchers and practitioners occasionally run into line search failure of the sort that we have described in \sectionref{sec:robust_line_search}.
We hope that our robust line search --- which requires only a few additional lines of code on top of the existing backtracking line search --- can eliminate the need for ad-hoc workarounds.

On that note, an experimental comparison to the line search method of Hager and Zhang [2005], which unlike ours requires the user to set an $\epsilon$ parameter, would be helpful.
However, the effort necessary to design a fair comparison is beyond the scope of our present study.

We have presented an in-depth experimental analysis of the convergence behavior of Projected Newton, demonstrating that this popular state-of-the-art method has severe drawbacks. It often performs considerably worse than Newton's method, which it is intended to replace.
Sometimes it needs more than an order of magnitude more iterations like we saw for the Armadillo Slingshot scenario.
Our findings may have significant implications for both past and future research.
PN is routinely used as a baseline for comparison.
Given its often suboptimal performance compared to Newton's method, the conclusions of some previous works may therefore warrant a new reading.
In the other direction, we hope that researchers will consider a variant of POD-Newton in place of PN for new projects, which may significantly accelerate convergence for a variety of problems.

We conclude by saying that POD-Newton and Kinetic Newton are both based on heuristic decision-making schemes.
We hope that our work will inspire future researchers to design further improved methods.

\section*{Acknowledgements}
We thank Lukas Westhofen for proofreading and Zachary Ferguson for assistance with the IPC Toolkit.
The Armadillo model is courtesy of Stanford Computer Graphics Laboratory.
This paper was funded by the Deutsche Forschungsgemeinschaft (DFG, German Research Foundation) --- Project number BE 5132/5-1.

\bibliography{bibliography}

\appendix
\section{Mathematical results}

\subsection{Positive definiteness of regularized Hessian}
\label{sec:pos_def_regularized_Hessian}
Let $\| \vec y \|_M^2 = \vec y^T \vec M \vec y$ be the squared norm of $\vec y$ induced by a symmetric positive definite matrix $\vec M$ (not necessarily the mass matrix) and let $\vec H$ be the Hessian matrix indicated in the regularized subproblem \eqref{eq:newton_regularized_subproblem}. Then the Hessian matrix of the regularized subproblem is $\vec H + \gamma_k \vec M$ and satisfies
\begin{align}
\begin{aligned}
\vec y^T (\vec H + \gamma_k \vec M) \vec y
= (\vec M^{\frac{1}{2}} \vec y)^T( \vec M^{-\frac{1}{2}} \vec H \vec M^{-\frac{1}{2}} + \gamma_k \vec I) (\vec M^{\frac{1}{2}} \vec y).
\end{aligned}
\end{align}
We define $\vec z = \vec M^{\frac{1}{2}} \vec y$ and define the symmetric eigendecomposition $\vec M^{-\frac{1}{2}} \vec H \vec M^{-\frac{1}{2}} = \vec U \vec \Sigma \vec U^T$, so that for $\vec y \neq \vec 0$ and sufficiently large $\gamma_k$,
\begin{align}
\vec y^T (\vec H + \gamma_k \vec M) \vec y
= (\vec U^T \vec z)^T (\vec \Sigma + \gamma_k \vec I) (\vec U^T \vec z) > 0,
\end{align}
and so the regularized Hessian is positive definite.

\subsection{Decay rates for robust line search}
\label{sec:robust_line_search_decay_rates}
Expanding terms evaluated at $\vec u^k + \alpha \Delta \vec u$ with Taylor series, we obtain
\begin{align}
\begin{aligned}
|\varepsilon_\text{est}| &= \frac{\alpha}{2} \Big| \Delta \vec u \cdot \Big( \nabla E(\vec u^k) - \nabla E(\vec u^k + \alpha \Delta \vec u) \Big) \Big|
\\
&= \frac{\alpha}{2} \Big| \Delta \vec u \cdot \Big( -\alpha \vec H(\vec u^k) \Delta \vec u + \mathcal{O}\Big(|\alpha \Delta \vec u|^2\Big) \Big) \Big|
\\
&= \mathcal{O} \Big( |\alpha \Delta \vec u|^2 \Big).
\end{aligned}
\end{align}
In contrast,
\begin{align}
\begin{aligned}
\Delta E_\text{approx} &= \frac{\alpha}{2} \Delta \vec u \cdot \Big( 2 \nabla E(\vec u^k) + \alpha H(\vec u^k) \Delta \vec u + \mathcal{O}(|\alpha \Delta \vec u|^2) \Big)
\\
&= \mathcal{O}(|\alpha \Delta \vec u |).
\end{aligned}
\end{align}

\subsection{Newton step is a first-order error estimate}
\label{sec:newton_first_order_estimate}
Assume that $E$ is sufficiently smooth, and let $\vec u^*$ be a nearby exact solution for the current solution estimate $\vec u$. Under the assumption that $\vec H(\vec u)$ is invertible and bounded in a region of $\vec u^*$ and $\vec u$ is contained inside this region, we have that the Newton step satisfies
\begin{align}
\begin{aligned}
\Delta \vec u &= - \vec H^{-1}(\vec u) \vec r(\vec u)
\\
&= - \vec H^{-1}(\vec u) \big[ \vec r(\vec u) - \vec r(\vec u^*) \big]
\\
&= - \vec H^{-1}(\vec u) \big[ \vec r(\vec u) - \vec r(\vec u + \vec e) \big]
\\
&= - \vec H^{-1}(\vec u) \big[ - \vec H(\vec u) \vec e - \mathcal{O}(\|\vec e \|^2) \big]
\\
&= \vec e + \mathcal{O}(\|\vec e \|^2),
\end{aligned}
\end{align}
where we have used the identity $\vec r(\vec u^*) = 0$ and standard Taylor series expansion.

\section{Swinging Beam with the Stable Neo-Hookean model}
\label{sec:swinging_beam_stable_neo_hookean}
\begin{figure}[htb!]
    \begin{center}
        \includegraphics{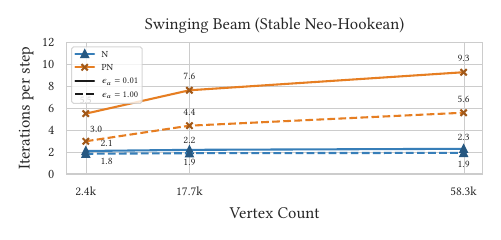}
    \end{center}
    \caption{Average number of solver iterations per time step for the \emph{swinging beam} experiment with the Stable Neo-Hookean material model.
    }
    \label{fig:swinging_beam_stable_neo_hookean}
\end{figure}
~
Our experiments in \sectionref{sec:projected_newton_convergence} use the Neo-Hookean material model, which sometimes presents numerical difficulties due to the strong nonlinearity in its logarithmic terms.
We repeat the experiment with the \emph{Stable Neo-Hookean} material model~\cite{SGK18}, a popular alternative that exhibits similar behavior but is inversion-safe and does not produce forces with magnitude that tend to infinity under compression.
The iteration counts for different resolutions of the beam are presented in \figref{fig:swinging_beam_stable_neo_hookean}.
The results are almost identical to the Neo-Hookean results from \figref{fig:teaser} (Right).
This is consistent with the characterization of Stable Neo-Hookean as an approximation of Neo-Hookean.
In this case, although deformations are large, they are not severe enough for the Stable Neo-Hookean approximation to deviate significantly from the true Neo-Hookean material.
Hence, since the material behavior is almost identical, we can expect the numerical behavior to be similar, as we have seen here.

\section{Boundary constraints}
\label{sec:implementation_boundary_constraints_details}
We consider two kinds of boundary constraints: \emph{direct imposition} and \emph{penalty}. We only consider simple constraints in the sense that some nodes are fully constrained and all others are free, i.e. $\vec u = (\vec u_F, \vec u_C)$, where $\vec u_F$ denotes the displacements of the free nodes and $\vec u_C$ the displacements of the constrained nodes.

\subsection{Direct imposition}
By directly imposing the constraints on the incremental potential \eqref{eq:backward_euler_incremental_potential}, the optimization problem becomes
\begin{align*}
\min_{\vec u_F} E(\vec u_F; \vec u_C).
\end{align*}
In our implementation, we re-use the assembly results of the unconstrained problem and modify gradients and Hessians to accommodate the constraints.

Direct imposition is problematic for \emph{moving} boundaries, since it is effectively \emph{teleporting} the constrained vertices, possibly moving the system into an invalid state with infinite energy, or alternatively a configuration which is hard for the solver to escape from.

\subsection{Penalty}
A simple alternative to direct imposition is using sufficiently stiff penalties.
There are many ways to formulate a penalty method.
We approximately enforce constraints with the penalty method by considering the modified incremental potential
\begin{align}
\min_{\vec u} E(\vec u) + \frac{\sigma}{2} \sum_{i \in \mathcal{C}} M_{ii} (u_i - u_{C,i})^2,
\end{align}
where $\mathcal{C}$ is the index set of constrained degrees of freedom and $\sigma > 0$ is a sufficiently large penalty factor.
By scaling with the mass matrix diagonal, we account approximately for the volume of each node, which makes the behavior of the constraint relatively consistent across different meshes and resolutions.

\end{document}